\documentclass[review,3p,times,authoryear]{elsarticle}

\usepackage{amssymb}
\usepackage{amsmath}

\usepackage{amsmath,amssymb}  
\usepackage{wasysym}          

\usepackage{float}
\usepackage{array}
\usepackage{booktabs}
\usepackage{multirow}
\usepackage{xcolor} 

\usepackage[most]{tcolorbox}
\usepackage{xcolor}

\usepackage{booktabs,makecell,array}
\renewcommand{\arraystretch}{1.05}

\definecolor{promptboxbg}{RGB}{248, 249, 250}      
\definecolor{promptboxframe}{RGB}{222, 226, 230}   
\definecolor{promptheading}{RGB}{0, 86, 179}       
\definecolor{placeholder}{RGB}{192, 80, 0}       

\newcommand{\promptvar}[1]{\textcolor{placeholder}{\texttt{\{#1\}}}}

\usepackage{hyperref}

\usepackage{booktabs}
\usepackage{enumitem}

\usepackage{listings} 
\tcbuselibrary{skins,breakable}

\lstdefinelanguage{CustomJava}[]{Java}{
	morekeywords={var}, 
}

\usepackage{multirow}

\journal{Computers in Industry}

\begin{document}

\begin{frontmatter}



\title{Synergistic Enhancement of Requirement-to-Code Traceability: A Framework Combining Large Language Model based Data Augmentation and an Advanced Encoder}




\author[hznu]{Jianzhang Zhang}
\ead{zjzhang@hznu.edu.cn}

\author[hznu]{Jialong Zhou}
\ead{jialongzhouzj@gmail.com}

\author[unf]{Nan Niu}
\ead{nan.niu@unf.edu}

\author[jx]{Jinping Hua}
\ead{huajinpin@foxmail.com}

\author[hznu]{Chuang Liu\corref{cor1}}
\ead{liuchuang@hznu.edu.cn}

\cortext[cor1]{Corresponding author.}

\affiliation[hznu]{
	organization={Department of Management Science and Engineering, Hangzhou Normal University},
	city={Hangzhou},
	state={Zhejiang},
	country={P.R. China}
}

\affiliation[unf]{
	organization={School of Computing, University of North Florida},
	city={Jacksonville},
	state={FL},
	country={USA}
}

\affiliation[jx]{
	organization={Jiangxi Provincial Institute of Cyber Security},
	city={Nanchang},
	state={Jiangxi},
	country={P.R. China}
}

\begin{abstract}
Automated requirement-to-code traceability link recovery, essential for industrial system quality and safety, is critically hindered by the scarcity of labeled data. To address this bottleneck, this paper proposes and validates a synergistic framework that integrates large language model (LLM)-driven data augmentation with an advanced encoder. We first demonstrate that data augmentation, optimized through a systematic evaluation of bi-directional and zero/few-shot prompting strategies, is highly effective, while the choice among leading LLMs is not a significant performance factor. Building on the augmented data, we further enhance an established, state-of-the-art pre-trained language model based method by incorporating an encoder distinguished by a broader pre-training corpus and an extended context window. Our experiments on four public datasets quantify the distinct contributions of our framework's components: on its own, data augmentation consistently improves the baseline method, providing substantial performance gains of up to 26.66\%; incorporating the advanced encoder provides an additional lift of 2.21\% to 11.25\%. This synergy culminates in a fully optimized framework with maximum gains of up to 28.59\% on $F_1$ score and 28.9\% on $F_2$ score over the established baseline, decisively outperforming ten established baselines from three dominant paradigms. This work contributes a pragmatic and scalable methodology to overcome the data scarcity bottleneck, paving the way for broader industrial adoption of data-driven requirement-to-code traceability.






\end{abstract}


\begin{highlights}
\item A synergistic framework is proposed to overcome the data scarcity bottleneck in requirements traceability by integrating LLM-driven data augmentation with an advanced encoder.
\item The effectiveness of LLM-driven data augmentation is systematically validated, revealing that the prompt engineering strategy is the dominant performance factor, not the choice of a specific LLM.
\item On its own, data augmentation yields dramatic performance gains of up to 26.66\% for an established state-of-the-art method, while an advanced, aligned pre-trained encoder offers an additional lift of up to 11.25\%.
\item The fully optimized framework decisively outperforms 10 baselines and achieves maximum gains of up to 28.59\% on $F_1$ score and 28.9\% on $F_2$ score over the established baseline method.
\end{highlights}


\end{frontmatter}




\section{Introduction}
\label{sec1}

In modern complex industrial systems, such as those in the aerospace and telecommunication sectors, requirements traceability (RT) is indispensable for ensuring quality and safety, particularly in safety-critical applications~\citep{Guo2025}. Formally, RT is the ability to link a requirement backward to its origins and forward to corresponding artifacts such as design specifications, source code, and test cases~\citep{Antoniol2025}. These links are fundamental for verifying that the final product precisely fulfills customer needs~\citep{Alturayeif2025}. However, these critical links to various downstream artifacts frequently go missing or become outdated during development, necessitating their systematic recovery. The field dedicated to this challenge is known as requirements traceability link recovery (RTLR). Failure to recover these links can lead to severe consequences, including significant quality degradation, economic losses, and safety hazards~\citep{Ramesh2001}, thereby derailing quality initiatives in large-scale industrial projects~\citep{Fucci2022}. Consequently, effective link recovery is not only a key enabler for essential software engineering tasks like change impact analysis and verification~\citep{Tian2021,Oosten2023}, but also a frequent mandate by regulatory standards to ensure compliance and manage risk~\citep{Wang2020,Mucha2024}. 

\begin{figure}[t]
	\centering
	\includegraphics[width=0.65\linewidth]{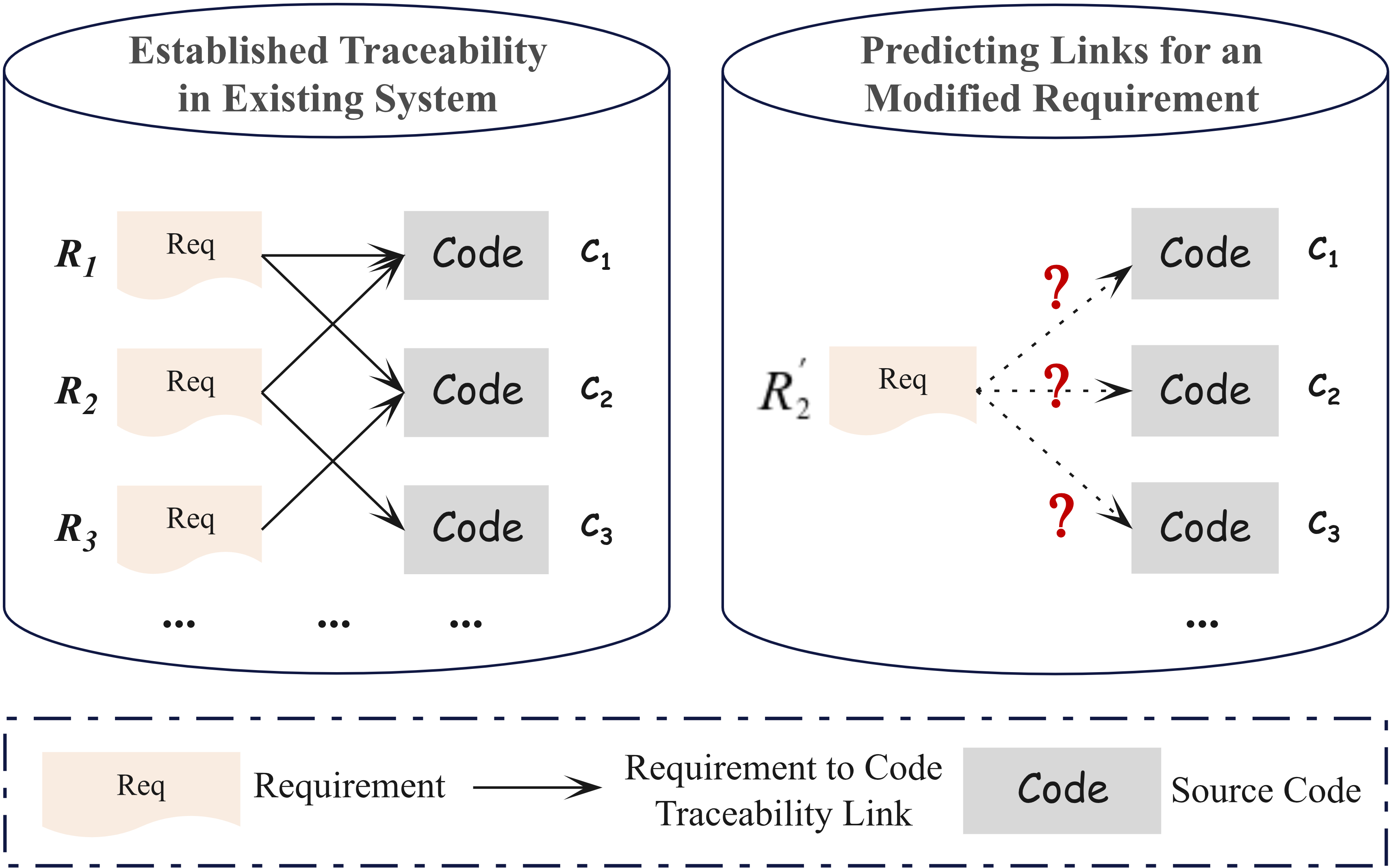}
	\caption{A Prototypical Scenario of Requirement-to-Code Traceability Link Recovery. The figure illustrates the recovery of links for a modified requirement ($R_{2}^{\prime}$), which is representative of broader RTLR challenges, such as handling new requirements and code artifacts.}\label{fig:RTLR}
\end{figure}

Within the broad domain of RTLR, recovering links between natural language requirements and their implementation in source code presents a particularly vital and persistent challenge, creating a significant conceptual gap between problem description and technical solution~\citep{Antoniol2002,Wan2025}. Therefore, this study focuses specifically on advancing RTLR techniques for requirement-to-code traceability. Figure~\ref{fig:RTLR} illustrates this challenge through a prototypical example. In the scenario depicted, an existing requirement, $R_2$, is modified, resulting in a new version, $R_{2}^{\prime}$. The original traceability links may no longer be valid, and the new, correct links for $R_{2}^{\prime}$ are unknown and must be recovered from the existing codebase. This situation is common during software maintenance and evolution. While this figure highlights the case of a modified requirement, the scope of RTLR is broader. The need for link recovery also arises in other critical scenarios, such as:

\begin{itemize}
	\item When entirely new requirements are introduced into the system.
	
	\item When new code artifacts are developed to implement existing requirements.
	
\item 	When existing code is refactored, split, or merged, thus altering its implementation of the original requirements.
\end{itemize}

Automated approaches to RTLR have historically followed two main paradigms: information retrieval (IR) and machine learning (ML). IR-based methods frame link recovery as a search problem~\citep{Wan2025}, employing statistical methods such as vector space models to measure textual similarity between software artifacts~\citep{Asuncion2010,Mahmoud2015,Gao2022,Li2025}. However, these methods are inherently limited by the vocabulary mismatch problem—the semantic gap between the distinct terminologies used in natural language requirements and programming language code. To overcome this limitation, the ML-based paradigm reframes RTLR as a predictive modeling problem, typically as a classification or ranking task~\citep{Alturayeif2025}. Instead of relying on direct textual similarity, ML-based methods learn to identify traceability links by discovering latent semantic patterns from labeled training data. This data-driven approach allows them to bridge the semantic gap by effectively handling challenges like synonymy and terminological ambiguity, where IR methods often fall short. The sophistication of ML techniques for RTLR has advanced significantly, progressing from probabilistic models~\citep{Moran2020} to deep learning architectures~\citep{Guo2017} and, more recently, pre-trained language models (PLMs)~\citep{Lin2021,Deng2024}. Despite their potential, the performance of these advanced models is often constrained by a critical bottleneck: the scarcity of large-scale, high-quality labeled datasets for training~\citep{Zogaan2017,Dong2022,Zhu2022,Lin2022}.

However, the exceptional performance of these models, from deep learning architectures to state-of-the-art pre-trained language models, is fundamentally predicated on the availability of large-scale, high-quality labeled data for training. In industrial practice, the manual curation and maintenance of traceability links from requirements to code is a notoriously labor-intensive, tedious, and often prohibitively expensive endeavor~\citep{Fucci2022}. Consequently, a severe scarcity of adequate training data is a pervasive challenge across most projects. This data-centric bottleneck emerges as the principal impediment to the scalable application of advanced ML techniques in industrial requirements traceability scenarios, thereby forming a critical research gap that demands innovative solutions.

To address this data scarcity bottleneck, this paper proposes and validates a synergistic framework for requirement-to-code traceability that integrates LLM-driven data augmentation with an advanced, aligned encoder. The principal contributions of this work are as follows:

\begin{itemize}
	\item \textbf{First, we establish a systematic methodology for LLM-driven data augmentation in RTLR.} Our comprehensive evaluation reveals that the prompt engineering strategy is the dominant performance factor, not the choice of a specific LLM, providing a practical guide for leveraging LLMs in this context.
	
	\item \textbf{Second, we quantify the powerful synergy between data augmentation and model architecture.} We demonstrate that data augmentation alone provides substantial performance gains for an established, state-of-the-art PLM based method. Furthermore, we show that incorporating a more advanced and aligned encoder yields a significant additional performance lift.
	
	\item \textbf{Finally, we deliver a new high-performance benchmark for requirement-to-code traceability.} The fully optimized framework achieves substantial performance gains, with improvements of up to 28.59\% on $F_1$ score and 28.9\% on $F_2$ score over the established baseline, and decisively outperforms a comprehensive suite of ten baselines from three dominant paradigms.
\end{itemize}

The remainder of this paper is organized as follows. Section 2 reviews related work. Section 3 details our proposed framework. Section 4 describes the experimental setup. Section 5 presents and analyzes the experimental results. Section 6 discusses the findings. Finally, Section 7 concludes the paper and outlines future research directions.

\section{Related Work}

Research in automated RTLR has evolved through three primary classes of techniques: information retrieval, traditional machine learning, and most recently, deep learning ~\citep{Antoniol2025,Alturayeif2025,Wan2025}. This section reviews representative studies from previous research, analyzing their contributions and persistent limitations to precisely situate the research gap that motivates our study.

\subsection{Traditional Approaches to RTLR}

Early automated approaches to RTLR primarily followed two technical paradigms: information retrieval and traditional machine learning.

Information retrieval methods, the pioneering techniques in automated traceability, frame the link recovery task as a search problem. These methods identify potential links by computing the textual similarity between artifacts~\citep{Mahmoud2015}, such as requirements and source code. Representative techniques include the vector space model (VSM)~\citep{Hayes2006,Mahmoud2014,Ali2019,Kuang2019,Wang2021b,Wang2022a}, latent semantic indexing (LSI)~\citep{Rahimi2018,Gao2022,Gao2022a,Lapena2022,Gao2024}, and latent dirichlet allocation (LDA)~\citep{Asuncion2010,Ali2015}. However, despite their widespread adoption due to simplicity and interpretability, the performance of these methods faces a fundamental ceiling known as the ``vocabulary mismatch problem"~\citep{Wan2025}. This issue arises from the significant semantic gap between the natural language vocabulary used in high-level requirement descriptions and the technical, programming language-specific terminology used in source code. By relying heavily on lexical similarity, IR methods struggle to effectively handle synonymy, terminological ambiguity, and domain-specific expressions, thereby limiting their efficacy in complex traceability tasks.

To overcome the limitations of IR, subsequent research turned to traditional machine learning methods~\citep{Mills2018}, reframing RTLR as a supervised classification problem. Instead of directly computing textual similarity, these models learn latent semantic patterns from labeled data to predict the existence of a link. Commonly employed classifiers include support vector machines (SVM)~\citep{Li2015,Ye2016,Xie2019}, k-nearest neighbors (KNN)~\citep{Mills2018,Wang2023}, logistic regression (LR)~\citep{Mazrae2021}, and random forests~\citep{Oosten2023,Dong2022}. However, the effectiveness of these traditional ML methods is constrained by two core limitations. First, their performance is highly dependent on the quality of feature engineering~\citep{Alturayeif2025}. Researchers must manually design and extract a variety of features, such as similarity scores from VSM or LSI and query quality metrics~\citep{Du2020}, to feed the model. Second, as supervised learning methods, they require a substantial amount of high-quality labeled data for training, a bottleneck often referred to as data scarcity~\citep{Dong2022,Zhang2023a}. In most practical projects, obtaining such a dataset is prohibitively expensive, which critically limits the application and performance of these methods.

\subsection{Deep Learning-based RTLR}

The advent of deep learning marked a significant evolution in RTLR, offering more powerful models to bridge the semantic gap~\citep{Wang2022}. Early DL-based approaches primarily utilized recurrent neural network (RNN) architectures, such as gated recurrent units (GRU) and long short-term memory (LSTM) networks~\citep{Guo2017,Xie2019}. These models, with their inherent capacity for sequence modeling, proved more effective at capturing the contextual nuances within and between software artifacts compared to their traditional ML predecessors. A cornerstone study by Guo et al.~\citep{Guo2017} demonstrated that RNN-based architectures could effectively learn semantic representations from artifacts to enhance traceability performance. Alongside RNNs, other architectures such as convolutional neural networks and ensemble methods like deep forest have also been explored to learn traceability patterns~\citep{Dai2023,Wang2023a}. However, a major limitation of these DL models is their ``data-hungry" nature. Typically trained ``from scratch", they require vast amounts of high-quality labeled data to learn meaningful patterns and avoid overfitting, a requirement that is seldom met in practice due to the high cost of manual annotation~\citep{Oosten2023,Zhu2022}.


A revolutionary breakthrough in the field arrived with the ``pre-train, fine-tune" paradigm, led by pre-trained language models such as BERT~\citep{Devlin2018}. Instead of learning from scratch, these models leverage knowledge acquired from massive, unlabeled text corpora, allowing them to achieve state-of-the-art performance with significantly less task-specific data. The pioneering work by Lin et al.~\citep{Lin2021} was the first to demonstrate the transformative potential of fine-tuning BERT-based models for RTLR, establishing that a simple architecture concatenating the requirement and code into a single input sequence far surpassed previous methods. Subsequent studies have further validated this paradigm, successfully applying various BERT-based models, including domain-specific variants like CodeBERT~\citep{Feng2020} and GraphCodeBERT~\citep{Guo2021}, to link diverse software artifacts such as issues and commits~\citep{Lin2022,Deng2024}. While PLMs dramatically improved data efficiency, their performance remains fundamentally tethered to the availability of sufficient in-domain labeled data for fine-tuning. Recent explorations also leverage LLMs directly, for instance through in-context learning via prompt engineering~\citep{Fuchss2025,Hassine2024}. However, consistently achieving high performance on complex, data-scarce tasks such as RTLR with these approaches remains an ongoing challenge. Consequently, data scarcity persists as the most critical bottleneck preventing the widespread industrial adoption of these advanced models, a gap our research directly aims to address.  


\subsection{Data Augmentation in Software Engineering}

Data augmentation is a widely recognized strategy for mitigating data scarcity, and its application has yielded significant success across various domains of software engineering (SE), such as software defect prediction~\citep{Mao2024}, source code learning~\citep{Dong2025}, and code summarization~\citep{Song2023}. Conventional approaches applied program transformations, method clustering, and mixing graph embeddings to augment source code datasets for tasks like bug detection, code commenting, and code summarization~\citep{Song2023,Yu2022,Dong2024}.

More recently, the capabilities of LLMs have been increasingly harnessed for data augmentation in SE tasks~\citep{Hou2024}. For example, LLMs have addressed class imbalance in task classification by generating synthetic descriptions via paraphrasing and creating variants from `pivot points' (e.g., different technologies or business domains)~\citep{Wysocki2025}. Similarly, for API mention recognition, LLMs have been used to generate diverse samples through targeted transformations like token replacement to address data shortages~\citep{Zhang2025}. LLM-based data augmentation also extends to creating novel training data. For instance, GPT-4o has generated natural language explanations for non-functional requirements, which were then used to fine-tune smaller, explainable models~\citep{Rejithkumar2025}.

\subsection{Positioning This Study}



The preceding review demonstrates that automated RTLR research, from early IR methods to modern DL models, has consistently grappled with challenges ranging from the ``vocabulary mismatch problem" to ``data scarcity". Despite the widespread and successful application of data augmentation in other SE tasks, its systematic application to RTLR remains a notably under-explored area. While some prior work has touched upon related concepts through transfer learning~\citep{Lin2021,Deng2024,Lin2022}, the targeted use of modern LLMs to directly generate synthetic requirement-code pairs for data augmentation in RTLR has not been systematically investigated.

This study is positioned to directly address this gap. By building upon the established PLM-based architectural paradigm while directly tackling the well-documented data scarcity bottleneck that limits prior work, our research advances the field through a multi-faceted, systematic investigation. Specifically, this work delivers the first comprehensive comparison of LLM-driven data augmentation strategies for RTLR, establishes a rigorous basis for LLM selection independent of specific models, and validates a complete synergistic framework that demonstrably pushes performance beyond established baselines.


%
%

\section{Methodology}

\begin{figure}[t]
	\centering
	\includegraphics[width=\linewidth]{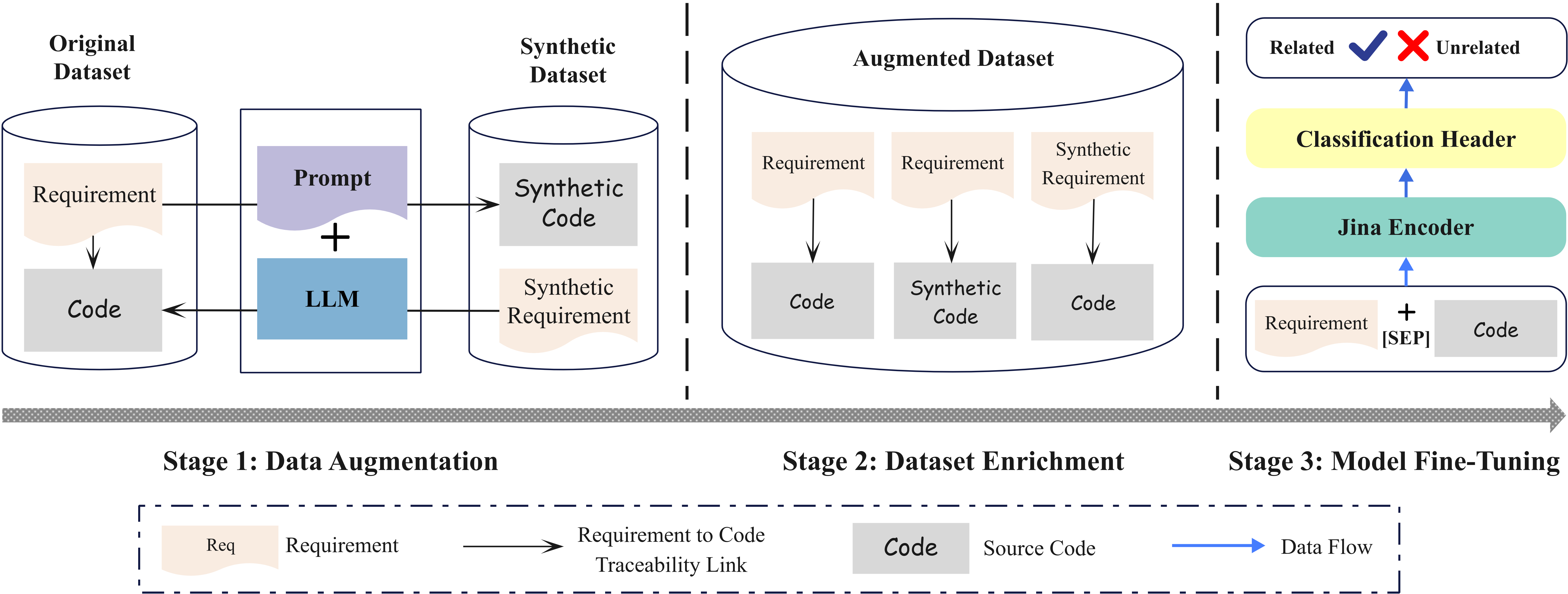}
	\caption{An Overview of the Three-Stage Synergistic Framework for Requirement-to-Code Traceability, Integrating LLM-driven Data Augmentation with an Enhanced Model Architecture.}\label{fig:method}
\end{figure}

To address the critical challenge of data scarcity in requirement-to-code traceability, we propose a comprehensive framework that synergistically combines LLM-driven data augmentation with an enhanced traceability model architecture. As illustrated in Figure~\ref{fig:method}, our methodology comprises three principal stages: Data Augmentation, Dataset Enrichment, and Model Fine-Tuning.

In the first stage, we systematically augment the original dataset, which consists of manually verified traceability links, by prompting LLMs to synthesize new code artifacts from existing requirements and, conversely, new requirement descriptions from existing code artifacts. In the second stage, this newly generated synthetic data is integrated with the original links to construct an enriched, augmented dataset, providing a more varied and robust training foundation. Finally, in the third stage, this augmented dataset is used to fine-tune a model built upon a pre-trained language model based architecture. This model employs an advanced encoder, Jina~\citep{Guenther2023}, which, in contrast to the baseline CodeBERT encoder~\citep{Feng2020}, is distinguished by a broader pre-training corpus and an extended context window, allowing it to better capture the complex semantic nuances between the artifacts. The subsequent subsections provide a detailed exposition of each stage.

\subsection{Data Augmentation}

%
%
%

\subsubsection{Bi-directional Generation Strategy}\label{sec:bi-gen}

The foundational stage of our framework is a systematic data augmentation process designed to address the critical scarcity of labeled training data. This process leverages the generative ability of LLMs~\citep{Zhao2023} to expand the original dataset into a more comprehensive and diverse training corpus. To ensure the quality and relevance of the synthetic data, we carefully use the existing ground-truth traceability links as high-fidelity seeds for generation.

Formally, let the original dataset be denoted as $D_{orig} = \{ (r_i, c_i) \}_{i=1}^{N}$, where each pair represents a verified traceability link between a requirement artifact $r_i$ and a code artifact $c_i$, and $N$ is the total number of links. We model the generation process via a generative function $\mathcal{G}(\cdot, P)$, which takes a source artifact and a prompt template $P$ as input. To maximize data diversity and bridge the semantic gap, we designed a bi-directional generation strategy.

This strategy operates along two complementary pathways:

\begin{enumerate}[label=(\arabic*)]
	\item \textbf{Requirement-to-Code (R-to-C) Synthesis}: For each pair $(r_i, c_i) \in D_{orig}$, the requirement $r_i$ is used as a seed. A synthetic code artifact, $c'_i$, is generated by the LLM, conditioned on the original requirement and a specific prompt template $P_{r \to c}$:
	\begin{equation}
		c'_i = \mathcal{G}(r_i, P_{r \to c})
	\end{equation}
	This process yields a new set of synthetic traceability links, $D_{r \to c} = \{ (r_i, c'_i) \}_{i=1}^{N}$.
	
	\item \textbf{Code-to-Requirement (C-to-R) Synthesis}: Conversely, for each pair $(r_i, c_i) \in D_{orig}$, the code artifact $c_i$ serves as the seed. The LLM is guided by a prompt template $P_{c \to r}$ to generate a synthetic natural language requirement, $r'_i$, that accurately describes the code's functionality:
	\begin{equation}
	r'_i = \mathcal{G}(c_i, P_{c \to r})
	\end{equation}
	This results in a second set of synthetic links, $D_{c \to r} = \{ (r'_i, c_i) \}_{i=1}^{N}$.
\end{enumerate}

Finally, the augmented training corpus, $D_{aug}$, is constructed by carefully forming the union of the original and the two newly generated synthetic datasets:
\begin{equation}
	D_{aug} = D_{orig} \cup D_{r \to c} \cup D_{c \to r}
\end{equation}

The rationale for this bi-directional approach is twofold. First, it is engineered to enhance data diversity by reflecting the inherent one-to-many relationships in software development. A single requirement $r_i$ can be realized through various functionally equivalent but stylistically distinct code implementations ($c_i, c'_i$, etc.). Our R-to-C synthesis path exposes the model to such implementation variance. Second, this strategy directly confronts the vocabulary mismatch problem. The functionality of a given code artifact $c_i$ can be described using different terminologies and phrasings ($r_i, r'_i$, etc.), which exists between high-level, user-centric descriptions and low-level technical artifacts~\citep{Wang2021a}. The C-to-R synthesis path generates these linguistic variations. By systematically creating data in both directions, we construct a richer, more robust training corpus that better prepares our traceability model for the complexities of real-world industrial scenarios.

\subsubsection{Prompt Templates}\label{sec:template}

To operationalize the bi-directional generation strategy detailed in Section~\ref{sec:bi-gen}, we systematically designed and formulated two distinct prompting paradigms: zero-shot and few-shot. This dual-strategy design was carefully chosen to comprehensively explore the capabilities of LLMs for this specialized task. The zero-shot method is intended to probe the model's intrinsic, pre-trained knowledge~\citep{Radford2019}, while the few-shot method investigates its capacity for in-context learning~\citep{Brown2020}. The objective of developing these parallel paradigms is to create a methodical foundation for identifying an effective prompt structure capable of generating high-quality, relevant synthetic data for RTLR.

\textbf{\textit{Zero-shot Prompts}}: The zero-shot method instructs the LLM to perform a task without providing any prior examples. The primary rationale for exploring this method is its scalability and efficiency, as it circumvents the need to select and curate specific examples for each generation instance. However, the performance of zero-shot prompting is highly sensitive to the clarity and completeness of the instructions. This sensitivity is a foundational challenge in prompt engineering, as LLMs require well-defined and unambiguous guidance to perform complex reasoning tasks effectively~\citep{Reimers2019}. To address this challenge and ensure precise, repeatable outputs, we engineered a highly structured prompt template. This prompt template is directly inspired by established prompt engineering frameworks, such as the CO-STAR model~\citep{GDSAD2023}, which advocates for providing explicit and comprehensive instructions to guide LLM behavior~\citep{GDSAD2023,Google2024}. The CO-STAR framework, an acronym for Context, Objective, Style, Tone, Audience, and Response, decomposes a prompt into six key components to minimize ambiguity and elicit higher-quality outputs. The role of each component is summarized in Table~\ref{tab:co-star-framework}.


\begin{table}[h!]
	\centering
	\caption{The CO-STAR Framework for Structured Prompt Design, based on the Prompt Engineering Playbook~\citep{GDSAD2023}.}
	\label{tab:co-star-framework}
	\renewcommand{\arraystretch}{1.2} 
	\begin{tabular}{l|p{0.75\linewidth}}
		\toprule
		\textbf{Component} & \textbf{Description} \\
		\midrule
		\textbf{Context (C)} & Provides background information and situational details necessary for the LLM to understand the task's environment~. \\
		\textbf{Objective (O)} & Clearly and specifically defines the primary task or goal the LLM is expected to accomplish~. \\
		\textbf{Style (S)} & Specifies the desired writing style for the output (e.g., academic, technical, mimicking a specific persona)~. \\
		\textbf{Tone (T)} & Defines the desired emotional or attitudinal tone of the response (e.g., formal, professional, empathetic)~. \\
		\textbf{Audience (A)} & Identifies the target audience for the generated output, guiding the LLM on appropriate vocabulary and complexity~. \\
		\textbf{Response (R)} & Explicitly constrains the desired output format, structure, and length (e.g., list, JSON, concise paragraph)~. \\
		\bottomrule
	\end{tabular}
\end{table}

Our zero-shot template is a direct implementation of this structured methodology. The purpose of this deliberate design is to provide the LLM with an unambiguous specification for the generation task, thereby maximizing control over the output and ensuring it aligns precisely with the stringent requirements of a training dataset. For instance, the \texttt{STYLE} and \texttt{AUDIENCE} components guide the LLM to mimic the distinct linguistic patterns of software engineers, while the \texttt{RESPONSE} component strictly governs the output format. As necessitated by our bi-directional strategy, two specific instantiations of this template were created: one for Requirement-to-Code synthesis and another for Code-to-Requirement synthesis, as detailed in Figure~\ref{fig:prompt-generate-code} and Figure~\ref{fig:prompt-generate-req} where \texttt{lang} denotes the programming language.

\begin{figure}[htpb]
	\centering
\begin{tcolorbox}[
	title=Zero-shot Prompt Template: Generate Code from Requirement,
	colback=promptboxbg,
	coltitle=black,
	colframe=promptboxframe,
	fonttitle=\bfseries,
	arc=3mm, 
	fontupper=\small
	]
	\textcolor{promptheading}{\textbf{CONTEXT}} \\
	I want to generate the corresponding \promptvar{lang} code based on the following requirements. \\
	\textit{\promptvar{requirements}}
	
	\textcolor{promptheading}{\textbf{OBJECTIVE}} \\
	Generate \promptvar{lang} code that fully implements the functions described in the requirements. The generated code must maintain high readability, completeness, accuracy, and compliance with \promptvar{lang} best practices.
	
	\textcolor{promptheading}{\textbf{STYLE}} \\
	Follow the writing style of a senior software development engineer implementing requirements.
	
	\textcolor{promptheading}{\textbf{TONE}} \\
	Accurate, clear, concise, readable, consistent, and reusable.
	
	\textcolor{promptheading}{\textbf{AUDIENCE}} \\
	The target audience for the \promptvar{lang} code includes other programmers, testers, code reviewers, and document writers. Tailor the \promptvar{lang} code to target what this audience typically looks for in software development products.
	
	\textcolor{promptheading}{\textbf{RESPONSE}} \\
	Provide only the \promptvar{lang} code. Ensure it exhibits the following qualities: clarity, conciseness, readability, modularity, maintainability, robustness, testability, efficiency, security, consistency, and scalability.
\end{tcolorbox}
    \caption{The Structured Zero-shot Prompt Template for Generating Code from a Requirement.}
\label{fig:prompt-generate-code}
\end{figure}

\begin{figure}[htpb]
	\centering
\begin{tcolorbox}[
	title=Zero-shot Prompt Template: Generate Requirement from Code,
	colback=promptboxbg,
	coltitle=black,
	colframe=promptboxframe,
	fonttitle=\bfseries,
	arc=3mm,
	fontupper=\small
	]
	\textcolor{promptheading}{\textbf{CONTEXT}} \\
	I want to summarize the corresponding requirements from the following code. \\
	\textit{\promptvar{code}}
	
	\textcolor{promptheading}{\textbf{OBJECTIVE}} \\
	Extract user requirements that focus on the goals users expect to achieve. Avoid involving code-level details; focus on user actions and expected results. Ensure the description is clear and accurately expresses the user's intention and expected experience.
	
	\textcolor{promptheading}{\textbf{STYLE}} \\
	Follow the writing style of a senior software engineer who defines requirements.
	
	\textcolor{promptheading}{\textbf{TONE}} \\
	Clear, accurate, concise, and formal.
	
	\textcolor{promptheading}{\textbf{AUDIENCE}} \\
	The target audience for the requirements includes quality assurance teams, testing engineers, business analysts, and development teams. Tailor the requirements to target what this audience typically looks for.
	
	\textcolor{promptheading}{\textbf{RESPONSE}} \\
	Provide only the requirement text. The text must be clear, internally consistent, and completely unambiguous.
\end{tcolorbox}
    \caption{The Structured Zero-shot Prompt Template for Generating a Requirement from Code.}
\label{fig:prompt-generate-req}
\end{figure}

\textbf{\textit{Few-shot Prompts}}: In contrast to the zero-shot method, the few-shot method is designed to leverage the in-context learning capabilities of LLMs by providing a concrete example of the desired input-output transformation~\citep{Brown2020}. The central hypothesis is that a well-chosen example can better anchor the LLM's understanding of the task, potentially leading to higher-quality and more consistent synthetic data.

Our few-shot prompt template was therefore systematically structured to consist of three parts: a concise task description, a single high-quality task example, and the seed artifact for generation. Unlike our zero-shot design, the more verbose CO-STAR framework was deliberately omitted from these templates for two strategic reasons. First, the primary directive in a few-shot setting is the example itself; a simpler task description ensures the LLM's focus is on replicating the provided pattern rather than interpreting complex instructions. Second, including a full input-output example significantly increases the prompt's token length, making a more concise instructional part necessary to stay within the context window limitations of the LLMs. Consistent with our bi-directional strategy, two parallel few-shot templates were developed to support both the R-to-C and C-to-R generation pathways, as detailed in Figure~\ref{fig:prompt-generate-code-few} and Figure~\ref{fig:prompt-generate-req-few}, respectively.

\begin{figure}[htpb]

	\centering
	\begin{tcolorbox}[
		title=Few-shot Prompt Template: Generate Code from Requirement,
		colback=promptboxbg,
		coltitle=black,
		colframe=promptboxframe,
		fonttitle=\bfseries,
		arc=3mm, 
		fontupper=\small
		]
		\textcolor{promptheading}{\textbf{TASK DESCRIPTION}} \\
		Generate the corresponding \promptvar{lang} code based on the following requirement. Generate \promptvar{lang} code for me and fully implement the functions described in the requirements. Must maintain high readability, completeness, accuracy, and compliance with \promptvar{lang} best practices. Give the \promptvar{lang} code without any other preamble text and requirements.\\
		\textit{\promptvar{requirements}}
		
		\textcolor{promptheading}{\textbf{EXAMPLE REQUIREMENTS}} \\
		\textit{\promptvar{example requirement}}
		
		\textcolor{promptheading}{\textbf{EXAMPLE OUTPUTS}} \\
		\textit{\promptvar{example code}}
	\end{tcolorbox}
	\caption{The Structured Few-shot Prompt Template for Generating Code from a Requirement.}
	\label{fig:prompt-generate-code-few}

\end{figure}

\begin{figure}[htpb]

	\centering
	\begin{tcolorbox}[
		title=Few-shot Prompt Template: Generate Requirement from Code,
		colback=promptboxbg,
		coltitle=black,
		colframe=promptboxframe,
		fonttitle=\bfseries,
		arc=3mm,
		fontupper=\small
		]
		\textcolor{promptheading}{\textbf{TASK DESCRIPTION}} \\
		Summarize the corresponding requirements from the following code. Extract user requirements that focus on the goals that users expect to achieve through this feature. Avoid involving code details and focus on user actions and expected results. Ensure that the description is clear and accurately expresses the user's intention and expected experience. Give the user requirements without any other preamble text and code. \\
		\textit{\promptvar{code}}
		
		\textcolor{promptheading}{\textbf{EXAMPLE CODE}} \\
		\textit{\promptvar{example code}}
		
		\textcolor{promptheading}{\textbf{EXAMPLE OUTPUTS}} \\
		\textit{\promptvar{example requirement}}
		
	\end{tcolorbox}
	\caption{The Structured Few-shot Prompt Template for Generating a Requirement from Code.}
	\label{fig:prompt-generate-req-few}

\end{figure}

\subsubsection{Synthetic Dataset Generation}

Generating the synthetic dataset encompasses two primary activities: instantiating the abstract prompt templates and using LLMs for generation.

\textbf{\textit{Prompt Instantiation}}: The abstract placeholders within the prompt templates (detailed in subsection~\ref{sec:template}) are systematically populated with specific content derived from the original dataset, specifically involving the following placeholders:
\begin{itemize}
	\item \texttt{\{lang\}}: This placeholder is replaced with the specific target programming language of the dataset being augmented (e.g., \textit{Java}, \textit{C\#}).
	\item \texttt{\{requirements\}} \& \texttt{\{code\}}: These are dynamically replaced with the requirement text and corresponding code text from each seed pair in the original dataset, serving as the core input for the generation task.
	\item \texttt{\{example requirements\}} \& \texttt{\{example code\}}: For the few-shot templates, these placeholders are populated with a single, fixed, high-quality example pair selected from the original dataset. This choice was made deliberately to maintain consistency across all few-shot generation instances for a given dataset, thereby minimizing variability introduced by the example itself.
\end{itemize}

\textbf{\textit{Generation Using LLMs}}: With the prompts instantiated, the actual generation of synthetic data is performed using a carefully selected panel of mainstream, closed-source LLMs. This selection was informed by recent comprehensive surveys of the field, which identify decoder-only architectures, e.g., the GPT series, as the dominant and most prevalently studied models in contemporary software engineering research~\citep{Hou2024}. For each seed pair from the original dataset, its corresponding instantiated prompt was provided to each LLM to generate a single synthetic artifact in one discrete API call. This one-to-one generation process was executed for every model, resulting in several parallel sets of synthetic data.

\textbf{\textit{Post-processing of Generated Data}}: To ensure the quality of the synthetic artifacts used in our augmented datasets, we performed a series of post-processing steps. For generated code, we utilized regular expressions to extract only the relevant code blocks, as LLM outputs were sometimes enclosed in explanatory text or formatting marks. Similarly, for generated requirements, we programmatically removed common introductory or summary phrases and any residual Markdown syntax to ensure a clean, consistent format for model training.

\subsection{Dataset Enrichment}

As depicted in Stage 2 of our framework (Figure~\ref{fig:method}), the final augmented dataset is constructed by systematically integrating the newly created synthetic data with the original dataset. The resulting corpus is thereby composed of three distinct types of requirement-to-code traceability pairs:
\begin{enumerate}[label=(\arabic*)]
	\item \textbf{Original Pairs}: Traceability links drawn directly from the original dataset, consisting of an original requirement and its corresponding original code artifact.
	\item \textbf{R-to-C Synthetic Pairs}: Links composed of an original requirement and the synthetic code generated for it by the LLM.
	\item \textbf{C-to-R Synthetic Pairs}: Links composed of a synthetic requirement generated by the LLM and its corresponding original code artifact.
\end{enumerate}

By combining these three components, the augmented dataset not only preserves the verified, ground-truth links but is also enriched with a diverse set of plausible, LLM-generated examples. This provides a substantially larger and more varied training foundation for the subsequent model fine-tuning stage. For a concrete illustration of this process, \ref{sec:data-examples} presents a seed pair from the iTrust dataset and its resulting synthetic versions.





\subsection{Model Fine-tuning}

The final stage of our framework, as delineated in Stage 3 of Figure~\ref{fig:method}, leverages the enriched dataset to train a robust traceability model capable of accurately classifying requirement-code pairs as either \textit{Related} or \textit{Unrelated}. This stage is predicated on two core components: a carefully selected model architecture and a systematic training regimen. The latter defines the specific learning objective and optimization algorithm used in the fine-tuning process.


\subsubsection{Architectural Design}

For the core of our traceability model, we employed a standard and empirically validated architecture for fine-tuning BERT-based models on classification tasks. Our rationale for this selection is threefold: its established effectiveness for general-purpose text classification~\citep{Sun2019}, its successful application to various software engineering tasks like bug and commit classification~\citep{VonderMosel2023}, and its proven superior accuracy for the specific challenge of requirement-to-code traceability compared to alternative designs~\citep{Lin2021}. This architecture processes each requirement-code pair through the following three sequential steps:

\begin{enumerate}[label=(\arabic*)]
\item \textbf{Input Formulation}: Given a requirement-code pair $(r, c)$, where $r = \{r_1, r_2, \dots, r_m\}$ and $c = \{c_1, c_2, \dots, c_n\}$ are sequences of tokens, they are concatenated into a single input sequence, $S$. This sequence is framed by special tokens \texttt{[CLS]} and \texttt{[SEP]} as follows:

\begin{equation}
	S = \text{[CLS]} \oplus r \oplus \text{[SEP]} \oplus c \oplus \text{[SEP]}
\end{equation}

where $\oplus$ denotes concatenation. This formulation enables the underlying BERT-based encoder to perform cross-attention between the requirement and code tokens within the same sequence. This allows the model to jointly learn the intricate semantic relationships and dependencies between the two distinct artifacts, which is fundamental for bridging the conceptual gap between them.

\item \textbf{Feature Extraction}: The input sequence $S$ is fed into the pre-trained encoder, $\mathcal{E}$, to produce a matrix of contextualized hidden states, $H \in \mathbb{R}^{L \times d_h}$, where $L$ is the sequence length and $d_h$ is the dimension of the hidden states:

\begin{equation}
	H = \mathcal{E}(S)
\end{equation}

\item \textbf{Classification}: An average pooling function, $\text{AvgPool}(\cdot)$, is applied over the hidden state matrix $H$ to derive a single, fixed-size feature vector $v \in \mathbb{R}^{d_h}$ that represents the aggregate semantics of the pair. This vector is then passed to a linear classification head with trainable parameters $(W, b)$ to compute the final probability distribution $p$ over two labels: \textit{Related} and \textit{Unrelated}.

\begin{equation}
v = \text{AvgPool}(H)
\end{equation}

\begin{equation}
p = \text{softmax}(Wv + b)	
\end{equation}

\end{enumerate}

\subsubsection{Training Regimen}

To effectively train the classifier, we designed a meticulous training regimen centered on a precise learning objective and a robust optimization algorithm.

%


\textbf{\textit{Loss Function}}: We frame the link recovery task as a binary classification problem. The model's objective is to learn a mapping that minimizes the divergence between its predicted probabilities and the ground-truth labels. To this end, we employ the binary cross-entropy loss function, a standard and theoretically sound choice for such tasks. 

For a single training sample, let $y$ be the ground-truth label (where $y=1$ for a \textit{Related} pair and $y=0$ for an \textit{Unrelated} pair) and $\hat{p}$ be the model's predicted probability for the \textit{Related} class. The loss $\mathcal{L}$ is then defined as:

\begin{equation}
	\mathcal{L}(\theta) = -[y \log(\hat{p}) + (1-y) \log(1-\hat{p})]
\end{equation}
where $\theta$ represents all trainable parameters of the model.

\textbf{\textit{Optimization Process}}: The goal of fine-tuning is to find the optimal set of parameters $\theta^*$ that minimizes the total loss across all samples in the training set $D_{\text{train}}$. We utilize the Adam optimizer~\citep{Kingma2015} to perform this optimization. Adam is a robust and widely adopted algorithm renowned for its efficiency and adaptive learning rate capabilities, making it particularly well-suited for fine-tuning PLM-based deep neural networks. During training, the model's parameters $\theta$ are iteratively updated via backpropagation to minimize the loss function, thereby effectively tuning the model to distinguish between genuine and spurious traceability links.

\section{Experimental Setup}

This section meticulously details the experimental design established to rigorously evaluate our proposed framework and systematically answer the research questions that guide our analysis in Section~\ref{sec:results}. We describe the benchmark datasets, the selection of encoders, the baseline methods, our training and implementation details, and the metrics chosen for performance evaluation. All experimental materials, including source code, datasets, and complete results, have been made publicly available to ensure full reproducibility\footnote{\url{https://figshare.com/s/db6b8b237d0fddfbdd30} (This link is currently anonymized for review purposes.)}.

\subsection{Datasets and Preprocessing}

Our evaluation is grounded in four publicly available and well-established benchmark datasets. These include iTrust, an open-source healthcare application for managing patient medical records; eTour, an electronic tourism guide developed by university students; the event-based traceability (EBT) system, a dynamic traceability infrastructure for artifact maintenance; and RETRO.NET, a tool for creating and maintaining traceability projects. Our selection was guided by a set of specific criteria, limiting our choice to systems that: (i) include trace links from requirements written in natural language to a code artifact, (ii) are documented and implemented in English, and (iii) represent non-trivial projects in terms of size. The iTrust, eTour, and EBT datasets~\footnote{https://gitlab.com/SEMERU-Code-Public/Data/icse20-comet-data-replication-package} are provided by the CoEST community, a trusted repository for traceability research, while the RETRO dataset\footnote{https://zenodo.org/records/1223649} was curated by Hayes et al.~\citep{Hayes2018}. The widespread recognition and frequent use of these particular datasets as benchmarks in the requirements traceability literature~\citep{Moran2020,Wang2024,Fuchss2025} further justify their suitability for this study. Furthermore, they represent a diversity of application domains and programming languages (Java, JavaScript, and C\#), which allows for a more robust evaluation of our framework's effectiveness. A statistical summary of these datasets is provided in Table~\ref{tab:datasets}.

\begin{table}[H]

	\caption{Statistical Summary of the Benchmark Datasets}
	\label{tab:datasets}
	\centering
	
	\begin{tabular}{lcccccc}
		\toprule
		\textbf{Dataset} &
		\makecell{\textbf{\# Req.}} &
		\makecell{\textbf{Avg. Req.}\\\textbf{Len. (Tokens)}} &
		\makecell{\textbf{\# Code}\\\textbf{Artifacts}} &
		\makecell{\textbf{Avg. Code}\\\textbf{Len. (Tokens)}} &
		\makecell{\textbf{\# Trace}\\\textbf{Links}} &
		\textbf{Language} \\
		\midrule

		\textbf{iTrust (Total)} & 195      &86.54       & 140       & 1994.76        & 402           & Java \& JavaScript \\
		\quad \textit{-- Java Part}      & 105    &-                  & 87          &-               & 286                     & Java         \\
		\quad \textit{-- JavaScript Part} & 90       &-                & 53           &-              & 116                     & JavaScript   \\
		
		\midrule

		\textbf{eTour}            & 58       &167.87                & 116             & 1537.06           & 308                     & Java         \\
		
		\textbf{EBT}              & 40    & 13.6                   & 50         & 484.71                & 98                      & Java         \\
		
		\textbf{RETRO.NET}        & 66     & 76.58                  & 78        &      2211.0            & 274                     & C\#          \\
		\bottomrule
	\end{tabular}
\end{table}

To prepare the data for our binary classification task, we performed two critical preprocessing steps: negative sampling and data splitting.

\textbf{\textit{Negative Sampling}}: We first addressed the fact that our augmented dataset, $D_{aug}$, contains exclusively positive links by constructing a balanced training corpus through a rigorous sampling procedure. For each requirement $r_i$ present in the augmented dataset, we identified the complete set of all code artifacts, denoted as $C_{linked}(r_i)$, to which it has a ground-truth traceability link. A negative sample $(r_{i},c'_{i})$ was then generated by pairing $r_i$ with a code artifact $c'_{i}$ randomly selected from the pool of all available code artifacts excluding those in $C_{linked}(r_i)$. This method explicitly ensures that any generated pair $(r_{i},c'_{i})$ is a true negative instance, critically avoiding the pitfall of incorrectly sampling a valid but different positive link. This process was carefully implemented to create a final dataset with a strict 1:1 ratio of positive to negative instances, which is essential for mitigating model bias~\citep{Lin2021}.

\textbf{\textit{Data Splitting}}: For model training and evaluation, the final dataset for each project was systematically partitioned into training (80\%), validation (10\%), and testing (10\%) sets. While both the training and validation sets incorporated synthetic data to enrich the learning process, it is crucial to note that the testing set was composed exclusively of original, manually verified data. This strict separation is fundamental to ensuring a rigorous and unbiased assessment of the model's generalization performance on unseen, real-world artifacts.


\subsection{LLMs}


For the data augmentation task, we utilized a panel of four leading, closed-source LLMs selected based on their state-of-the-art performance during our experimental window (Q2-Q3 2024). Their top-tier status at that time is corroborated by the LMArena Leaderboard, an established industry benchmark\footnote{\url{https://huggingface.co/spaces/lmarena-ai/lmarena-leaderboard/blob/main/elo_results_20240611.pkl}}. The specific models employed in our experiments were Gemini 1.5 Pro\footnote{\url{https://deepmind.google/technologies/gemini/}}, Claude 3 Opus\footnote{\url{https://www.anthropic.com/news/claude-3-family}}, GPT-3.5 Turbo\footnote{\url{https://openai.com/blog/new-models-and-developer-products-announced-at-devday}}, and GPT-4o\footnote{\url{https://openai.com/index/hello-gpt-4o/}}. By employing this diverse set of state-of-the-art models, we ensure that our findings regarding the effectiveness of data augmentation and prompting strategies are robust and not contingent on a single proprietary model or vendor.

\subsection{Model Encoders}

The encoder is the central component of our traceability model, tasked with transforming both natural language requirements and source code into meaningful semantic representations. To rigorously investigate how an encoder's characteristics influence performance, we selected two distinct BERT-based models for evaluation based on their differing language coverage and context window limitations.

\begin{itemize}
	\item \textbf{CodeBERT}: We selected CodeBERT~\citep{Feng2020} to serve as our baseline encoder. It is a foundational and widely adopted pre-trained model for tasks involving both natural and programming languages. It was trained on a large-scale corpus covering six programming languages (Python, Java, JavaScript, PHP, Ruby, and Go) and is limited to a maximum input sequence of 512 tokens. While C\# is not in its original training set, its proven generalization capabilities make it a robust starting point for our experiments.
	
	\item \textbf{Jina}: As our advanced encoder for comparison, we selected the Jina-v2 encoder~\citep{Guenther2023} to investigate whether broader language coverage and a longer context window enhance traceability performance. This model was deliberately chosen for its superior alignment with our task; it is pre-trained on a diverse set of over 30 programming languages, a corpus that explicitly includes all languages used in our experimental datasets (Java, JavaScript, and C\#). Furthermore, it supports a significantly longer input sequence of up to 8192 tokens due to the symmetric bidirectional variant of ALiBi~\citep{Press2022}. This generous context window is crucial, as it is sufficient to accommodate the full concatenated length of even the longest requirement-code pairs in our datasets, thereby eliminating information loss due to truncation—a key limitation of the baseline encoder.
	
	
\end{itemize}

This deliberate selection of encoders with contrasting characteristics allows us to directly assess the benefits of a more advanced and aligned encoder.

\subsection{Baselines for Comparison}\label{sec:baselines}

To comprehensively evaluate the effectiveness of our framework, we benchmark it against a wide array of established techniques for traceability link recovery~\citep{Alturayeif2025,Wan2025}. These baselines were carefully selected to represent three major classes of techniques in the field:


\begin{itemize}
	\item \textbf{Information Retrieval (IR)}: We include three classic IR methods: latent dirichlet allocation (LDA)~\citep{Asuncion2010}, latent semantic indexing (LSI)~\citep{Rahimi2018}, and the vector space model (VSM)~\citep{Hayes2006}.

\item \textbf{Machine Learning (ML)}: We selected three representative supervised learning classifiers~\citep{Mills2018}: k-nearest neighbors (KNN), logistic regression (LR), and a support vector machine (SVM).

\item \textbf{Deep Learning (DL)}: We implemented four sequential models, representative of the RNN-based architectures validated in the cornerstone study by Guo et al.~\citep{Guo2017}: gated recurrent unit (GRU), long short-term memory (LSTM), and their bidirectional variants (Bi-GRU and Bi-LSTM).

\textbf{PLM-based State-of-the-Art (SOTA)}: We include the seminal work by Lin et al.~\citep{Lin2021}, which represents the established SOTA in PLM-based traceability. Specifically, we implemented their best-performing ``Single" architecture, which utilizes CodeBERT as its underlying encoder to concatenate the requirement and code artifacts into a single input sequence. This model serves as our primary baseline, as our framework is designed to directly enhance this powerful and empirically validated paradigm.


\end{itemize}

For all baseline models, we strictly adhered to the hyperparameter configurations reported in their original respective publications to ensure a fair and faithful comparison.

\subsection{Training and Implementation Details}

This subsection details the configuration and hardware environment for our experiments.

\textbf{Stochastic Models (DL and PLM-based)}: To ensure the reliability and stability of our findings while mitigating the effects of stochasticity, all experiments involving the DL baselines, our proposed framework (using the Jina encoder), and the established SOTA baseline (using the CodeBERT encoder) were systematically repeated five times using distinct random seeds (2014-2018). The final results reported in this paper represent the average of these runs.

For the PLM-based models specifically, they were fine-tuned using a set of carefully selected hyperparameters based on established practices and preliminary tuning. The initial learning rate was set to $1 \times 10^{-5}$, the batch size was 8, and the training proceeded for 20 epochs. The maximum input sequence length was generally set to 512 tokens; to specifically assess the impact of input truncation, we also conducted experiments with an extended sequence length of 1024 tokens for the Jina encoder.

\textbf{Deterministic Baselines (IR and ML)}: For the IR and traditional ML baseline models, which are deterministic, we strictly adhered to the hyperparameter configurations reported in their original respective publications to ensure a fair and faithful comparison.

\textbf{Hardware Environment}: All experiments were conducted on a server equipped with two NVIDIA P100 GPUs, each with 16GB of memory.

%
%

\subsection{Evaluation Metrics}

To provide a comprehensive evaluation, we assess model performance using metrics that are widely adopted for evaluating traceability link recovery tasks~\citep{Guo2017,Lin2021,Wang2024}. As our framework models this task as a binary classification problem, we utilize the following standard metrics:

\begin{itemize}
	\item \textbf{Accuracy}: The overall proportion of correct classifications.
	\item \textbf{Precision}: The proportion of predicted positive links that are correct, measuring the model's exactness.
	\item \textbf{Recall}: The proportion of actual positive links that are correctly identified, measuring the model's completeness.
	\item \textbf{$F_1$ Score}: The harmonic mean of Precision and Recall, providing a balanced measure of overall performance.
	\item \textbf{$F_2$ Score}: A variant of the F-score that weighs Recall more heavily than Precision. We deliberately include this metric because, in the context of requirements traceability, failing to recover a true link (a false negative) is typically more costly than proposing an incorrect one (a false positive). The $F_2$ score thus offers critical insight into the model's practical utility in minimizing these costly errors~\citep{Berry2021}.
\end{itemize}

\section{Results and Analysis}\label{sec:results}

This section presents a systematic evaluation of our proposed framework. The overarching goal of this study is to improve the effectiveness of requirement-to-code traceability by synergistically combining LLM-based data augmentation with an advanced, aligned pre-trained encoder. To rigorously validate our approach and present the findings with clarity, we structure our analysis around four guiding research questions (RQs). The following subsections are each dedicated to one of these RQs, systematically dissecting our experimental findings. 

%

\subsection{Effectiveness of Data Augmentation via Prompting Strategies}\label{rq1}

\begin{figure*}[!ht]
	\centering
	\includegraphics[width=1\linewidth]{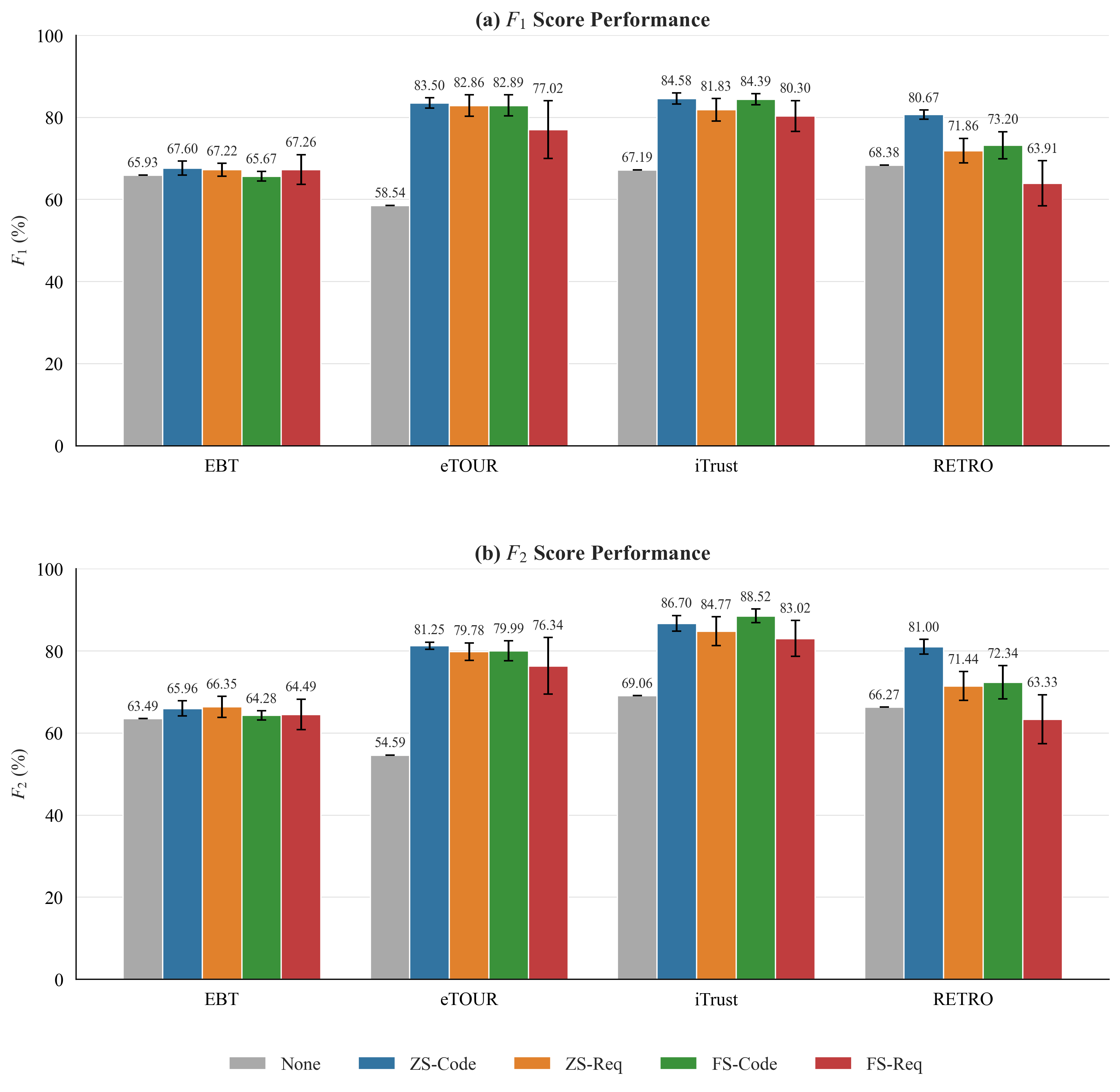}
	\caption{Performance Comparison of Four Prompting Strategies for Data Augmentation against a Non-augmented Baseline, Showing the (a) $F_1$ and (b) $F_2$ scores. Each bar represents the performance of a given configuration averaged across the four LLMs, with the error bars indicating the standard deviation. The `None' configuration serves as the non-augmented baseline. ZS-Code and ZS-Req denote the zero-shot code and requirement generation strategies, while FS-Code and FS-Req denote the few-shot strategies.}
	\label{fig:5m_4llm}
\end{figure*}

\textbf{\textit{Motivation and Method}}: Recognizing that the design of prompt templates plays a crucial role in shaping the quality and characteristics of synthetic data generated by LLMs, our initial research question (RQ1) is twofold: first, to determine if LLM-based data augmentation is an effective approach for improving traceability model performance, and second, if so, which prompting strategy is the most effective. To this end, we conducted a comprehensive experiment: for each of the four benchmark datasets, we generated synthetic data using four distinct prompting strategies across four leading LLMs. Each of these augmented datasets was then used to train our PLM-based baseline model (configured with the CodeBERT encoder) and benchmarked against a non-augmented baseline. To provide a clear and robust summary of these results, Figure~\ref{fig:5m_4llm} presents the performance of each strategy averaged across the four LLMs, with error bars indicating the standard deviation. For the comprehensive results of performance for each combination of LLM, prompting strategy, and evaluation metric, please refer to Table~\ref{tbl:combination1} and Table~\ref{tbl:combination2} in the~\ref{sec:comprehensive-results}.

\textbf{\textit{Findings and Interpretation}}: The results presented in Figure~\ref{fig:5m_4llm} unequivocally demonstrate that LLM-based data augmentation provides substantial performance improvements over the non-augmented baseline. These gains are observed across both the $F_1$ score (Figure~\ref{fig:5m_4llm}~(a)) and the recall-sensitive $F_2$ score (Figure~\ref{fig:5m_4llm}~(b)) and are most pronounced on the eTOUR dataset. On this dataset, for instance, the top-performing ZS-Code strategy boosted the average $F_1$ score by a remarkable 24.96 percentage points (from 58.54\% to 83.50\%), and, even more strikingly, the average $F_2$ score by 26.66 percentage points (from 54.59\% to 81.25\%). Similarly dramatic gains are also visible on the iTrust and RETRO.NET datasets. The effect was more modest on the EBT dataset, where $F_1$ scores improved by a smaller margin of 1.67 percentage points. We attribute this to the compounded data sparsity challenges unique to this dataset. As detailed in Table~\ref{tab:datasets}, EBT is by far the smallest of the four benchmarks, containing only 98 trace links. While our augmentation process adds one synthetic pair for each original link, the absolute number of newly generated examples remains the lowest among all datasets, limiting the overall learning benefit. This issue is further exacerbated by the extreme brevity of its artifacts; not only are its requirements the shortest (avg. 13.6 tokens), but its code artifacts are also significantly more concise (avg. 484.71 tokens) than those in other datasets. This lack of rich semantic context likely poses a dual challenge: it provides less material for the LLM to generate high-quality synthetic data, and less information for the downstream model to learn complex patterns from. It is also noteworthy that even on the RETRO.NET dataset, where the C\# language is outside CodeBERT's primary pre-training corpus, most augmentation strategies still yielded a discernible performance lift.

A closer analysis of the strategies reveals a clear pattern, identifying a top tier of three highly effective approaches: using zero-shot prompts to generate requirements, using zero-shot prompts to generate code, and using few-shot prompts to generate code. These three strategies consistently cluster at the top of the performance bar charts, with their relative effectiveness often fluctuating depending on the specific dataset; no single one proved to be universally superior. In contrast, the strategy of using few-shot prompts to generate requirements was notably less effective. This is best illustrated on the RETRO dataset, where this strategy achieved an $F_1$ score of only 63.91\%, underperforming the non-augmented baseline's score of 68.38\%. Furthermore, the small error bars on each bar, representing the standard deviation across the four LLMs, visually suggest that the performance variation among the different LLMs is minimal, a finding that will be rigorously evaluated in the next subsection (RQ2).

\textbf{\textit{Answer to RQ1}}: Our analysis confirms that LLM-based data augmentation is a highly effective method for this task, with the best approach being the selection of a prompting strategy from a top tier of three comparable options. This conclusion is based on two key findings: first, every data augmentation prompting strategy provides substantial performance improvements over the non-augmented baseline, confirming the overall effectiveness of the LLM-based data augmentation. Second, a clear performance pattern identified a high-performing group of three prompting strategies, i.e., zero-shot requirement and code generation, and few-shot code generation.



\subsection{Impact of LLM Choice on Data Augmentation Effectiveness}

\textbf{\textit{Motivation and Method}}: Having established in Section~\ref{rq1} that our data augmentation strategies can substantially improve model performance, our second research question (RQ2) logically follows: does the choice of a specific LLM for generation yield a significant performance difference? To answer this, we conducted a statistical analysis of the experimental results already presented in Figure~\ref{fig:5m_4llm}. For each prompting strategy, we compared the performance of our  PLM-based baseline model (still configured with the CodeBERT encoder) when trained on the four parallel augmented datasets, each created using synthetic data from one of the four leading LLMs: Gemini 1.5 Pro, Claude 3, GPT-3.5, and GPT-4o. To ascertain whether any observed performance differences were statistically significant, we employed the Wilcoxon signed-rank test, with a $p$-value threshold of $< 0.05$. To concisely present the results, Table~\ref{tab:p_value_summary} summarizes the findings. For each dataset, it shows the minimum and maximum $p$-values observed across the full spectrum of 20 comparison scenarios (5 evaluation metrics × 4 prompting strategies).


\begin{table}[h!]
	\centering
	\caption{Summary of $p$-value Ranges from Pairwise LLM Comparisons Across All Datasets. The range reflects the minimum and maximum $p$-values observed across all metrics and prompting strategies for each dataset.}
	\label{tab:p_value_summary}
	\begin{tabular}{lc}
		\toprule
		\textbf{Dataset} & \textbf{Range of $p$-values (min -- max)} \\
		\midrule
		EBT        & [$0.059$ -- $1.0$] \\
		eTOUR      & [$0.063$ -- $1.0$] \\
		iTrust     & [$0.063$ -- $1.0$] \\
		RETRO.NET  & [$0.063$ -- $1.0$] \\
		\bottomrule
	\end{tabular}
\end{table}

\textbf{\textit{Findings and Interpretation}}: The results of our statistical analysis, summarized in Table~\ref{tab:p_value_summary}, provide an unequivocal and globally consistent finding. The table clearly shows that for every dataset, the minimum $p$-value observed across the full spectrum of 20 comparison scenarios (5 evaluation metrics × 4 prompting strategies) remained well above the 0.05 significance threshold. This lack of statistically significant difference is, in itself, a highly significant finding. It indicates that the effectiveness of our data augmentation framework is not tightly coupled to a specific, proprietary LLM. Rather, the success of the approach appears to be robust across a range of leading models. This finding suggests that the prompting strategy, as analyzed in RQ1, is the dominant factor in generating high-quality synthetic data, while the choice between these top-tier LLMs is a secondary concern. This decouples the solution from any single model provider, enhancing its generalizability and practical applicability for tackling the data scarcity problem in requirements traceability.

\textbf{\textit{Answer to RQ2}}: Our comparative analysis clearly shows no single LLM among the tested top-tier proprietary models demonstrated a statistically significant performance advantage. The central conclusion is that the success of our data augmentation approach is not contingent on a specific LLM, but rather on the prompting strategy employed. This finding effectively decouples our framework from any particular proprietary model provider, affirming its generalizability and practical value for the field.

\subsection{Impact of Encoder Selection and Configuration}\label{sec:rq3}

\textbf{\textit{Motivation and Method}}: The encoder forms the architectural core of our traceability model, tasked with transforming raw text (both natural and programming languages) into rich semantic representations. Our third research question (RQ3) explores two primary avenues for its optimization: (1) the impact of an encoder's pre-training on a broader and more relevant corpus of programming languages, and (2) the effect of extending the maximum input sequence length to mitigate information loss from truncation. To systematically investigate these two avenues, we designed the following two-part experiment.

\begin{figure*}[h!]
	\centering
	\includegraphics[width=1\linewidth]{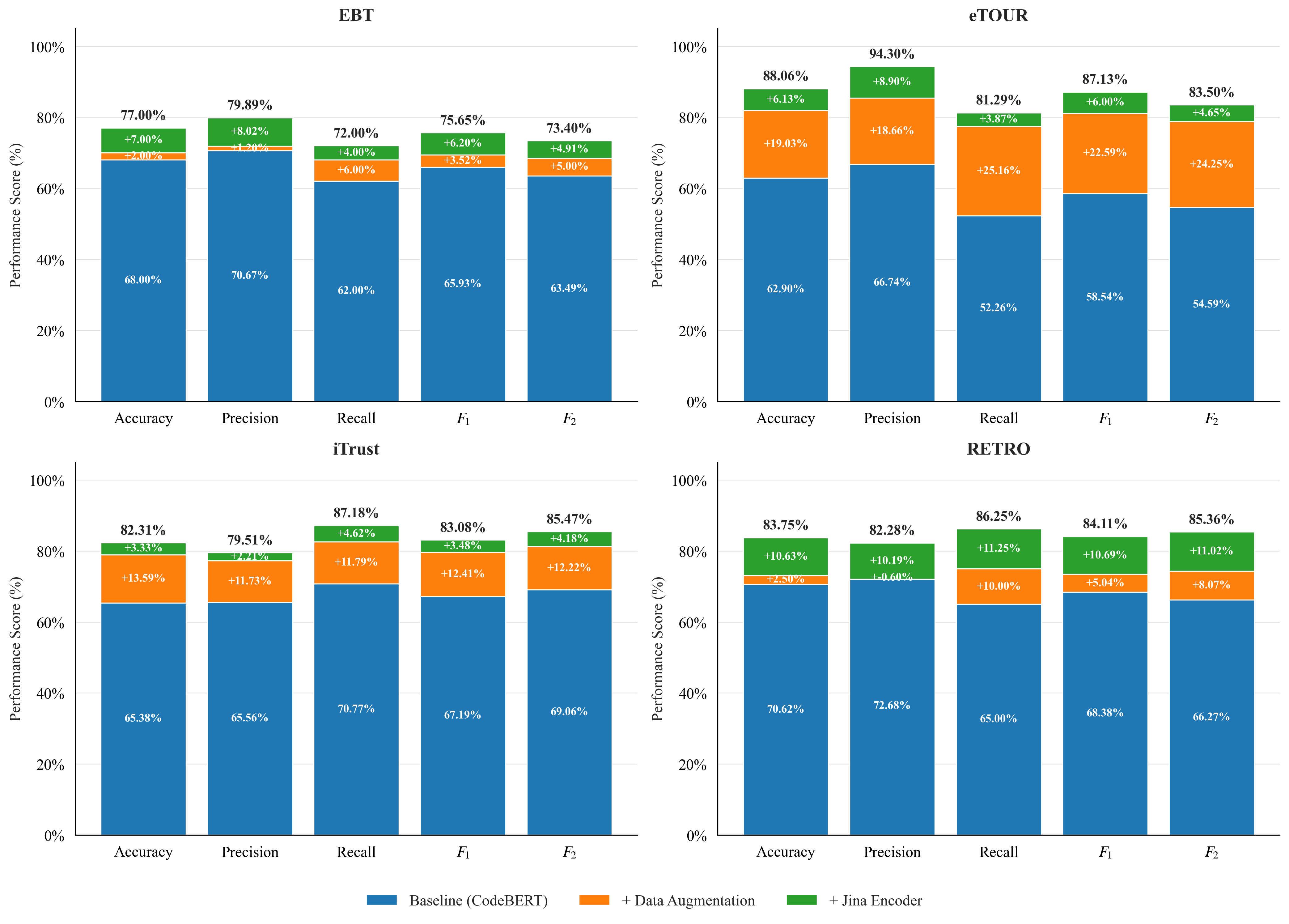}
	\caption{Comparative Analysis of the CodeBERT and Jina Encoders on Augmented Datasets. The chart compares the performance of the model using the Jina encoder against the CodeBERT encoder, with both operating on the same high-performing augmented dataset. The performance of the non-augmented CodeBERT model is also included as a baseline for contextual reference.}
	\label{fig:codeBERT_jina}
\end{figure*}

First, in an experiment designed to assess the impact of pre-training alignment, we compare the performance of our baseline CodeBERT encoder against Jina-v2, a model deliberately selected for its extensive pre-training on a corpus that includes all programming languages present in our datasets. This comparison was conducted under a controlled condition: we used an augmented dataset generated by GPT-4o employing the zero-shot requirement generation strategy. This strategy was chosen as a representative example from the top tier of high-performing strategies identified in RQ1, while the selection of GPT-4o was informed by the finding from RQ2 that all tested LLMs perform comparably. The results of this comparison are presented in the four subplots of Figure~\ref{fig:codeBERT_jina}. Each subplot uses stacked bar charts to visualize the incremental performance gains across the five metrics for three key configurations: the non-augmented CodeBERT baseline, the augmented CodeBERT model, and the augmented Jina model.

\begin{figure*}[h!]
	\centering
	\includegraphics[width=1\linewidth]{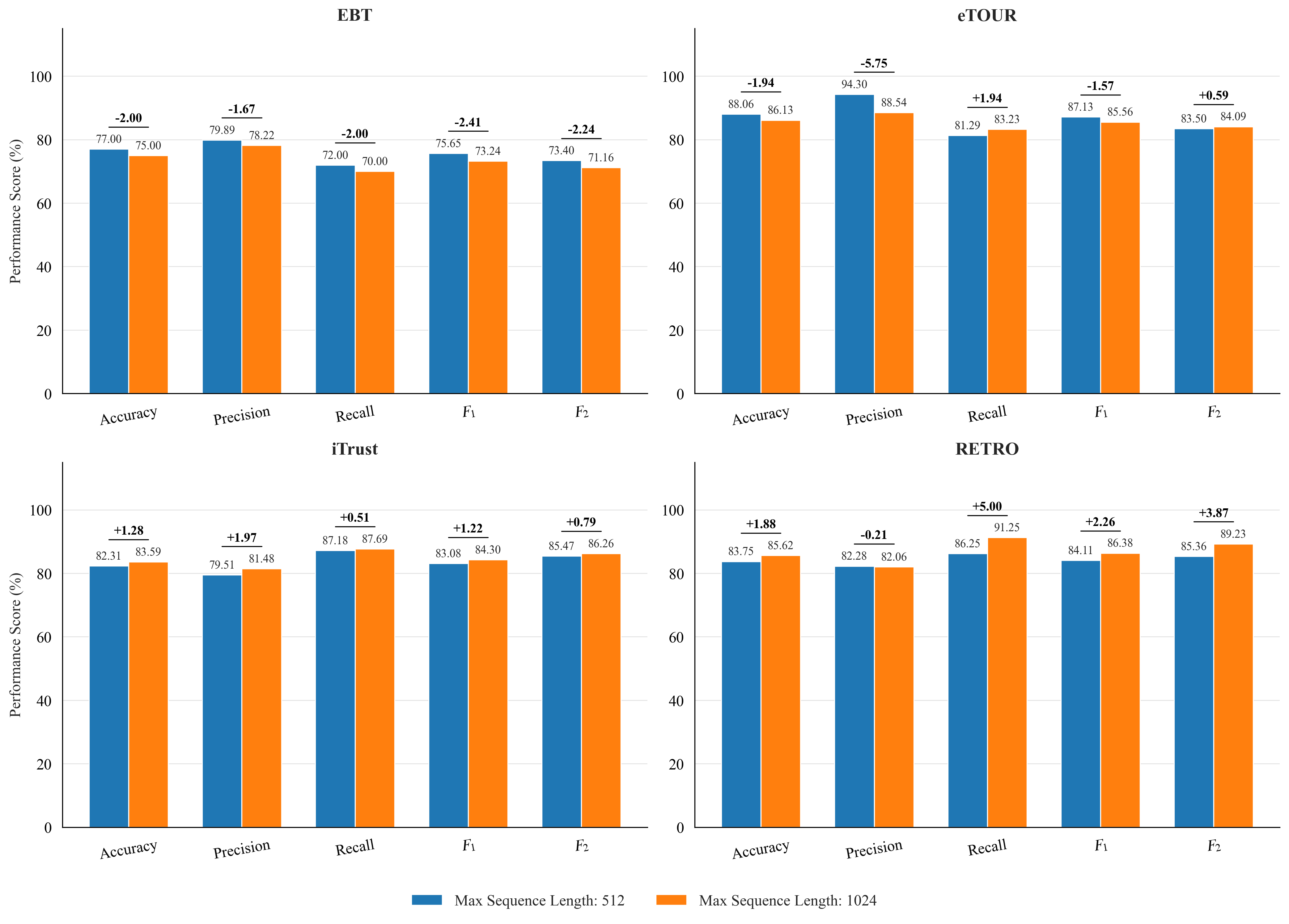}
	\caption{Evaluating the Impact of Input Sequence Length on the Jina Encoder's Performance. Each chart compares the model's performance when the maximum sequence length is set to 512 versus 1024 tokens on the same augmented dataset across the four benchmarks.}
	\label{fig:512_1024}
\end{figure*}

Second, in an experiment focused on the effect of input sequence length, we leverage the Jina-v2 encoder's capability to handle longer inputs and compare its performance with maximum sequence lengths set to 512 versus 1024 tokens. The results for this second experiment are displayed in Figure~\ref{fig:512_1024}, which also consists of four subplots, one for each dataset. Within each subplot, side-by-side bar charts are used to directly compare the performance of the two sequence length configurations across the five evaluation metrics.

\textbf{\textit{Findings and Interpretation}}: The results of our first experiment, presented in Figure~\ref{fig:codeBERT_jina}, are unequivocal: selecting an encoder with broader and more relevant language pre-training yields substantial performance gains. This is particularly insightful when considering the fundamental differences between the two encoders: while CodeBERT's pre-training corpus does not include C\#, the Jina encoder was deliberately chosen for its extensive training on over 30 programming languages, a set which explicitly includes all languages in our datasets (Java, JavaScript, and C\#). The tangible benefits of this superior pre-training are clearly visible across all datasets, such as on eTOUR where the Jina model achieved a final precision of 94.30\%. Notably, the advantage of Jina's broader language coverage is most powerfully demonstrated on the C\#-based RETRO.NET dataset. As illustrated in Figure~\ref{fig:codeBERT_jina}, this is precisely the scenario where replacing CodeBERT with Jina provides the largest incremental performance boost (e.g., an additional +11.25 percentage points on the recall). This result strongly indicates that when a baseline encoder lacks specific pre-training for a target language, it must rely on generalization, whereas an encoder like Jina, which has been explicitly trained on that language, provides a decisive performance advantage. This confirms that aligning an encoder's pre-training expertise with the target programming languages is a critical optimization for enhancing traceability performance.


The results of our second experiment reveals a nuanced, dataset-dependent pattern as shown in Figure~\ref{fig:512_1024}. An analysis of the datasets' artifact lengths, presented in Table~\ref{tab:datasets}, reveals a pattern that may help explain this divergence. We observe that the iTrust and RETRO datasets, which contain significantly longer code artifacts on average (mean of 1994.76 and 2211.0 tokens, respectively), both benefited from extending the sequence length from 512 to 1024 tokens. This observation supports the plausible interpretation that for projects with long artifacts that would otherwise be heavily truncated, a larger context window is beneficial as it reduces information loss. Conversely, for the EBT and eTOUR datasets, which feature more concise code artifacts on average (mean of 484.71 and 1537.06 tokens, respectively), the same extension led to a marginal but consistent decrease in performance. This could suggest that when source artifacts are short enough to fit within the standard window, a larger context may introduce non-essential context or noise, slightly hampering the model's focus. This finding indicates that the optimal sequence length is not a universal constant but a hyperparameter that should be empirically validated on a case-by-case basis. Simply defaulting to the largest possible sequence length is not a guaranteed strategy for performance enhancement.

\textbf{\textit{Answer to RQ3}}: Our experiments on encoder optimization lead to two key conclusions. We confirm that selecting an encoder whose pre-training corpus explicitly covers the target programming languages, such as Jina over CodeBERT, is a critical and consistently effective optimization strategy. In contrast, we find that increasing the maximum input sequence length is not a universally applicable enhancement. Its effect is dataset-dependent, highlighting that it should be treated as a project-specific hyperparameter to be tuned rather than a default improvement.

\subsection{Performance Comparison Against Baseline Methods}

\textbf{\textit{Motivation and Method}}: Having validated the effectiveness of data augmentation via prompting strategies (RQ1), confirmed the comparable performance of leading LLMs on data augmentation effectiveness (RQ2), and identified a superior pre-trained encoder (RQ3), our final research question (RQ4) is to evaluate the overall effectiveness of our fully optimized framework. The ultimate measure of our proposed approach's value is its performance relative to established techniques in the field.



\begin{figure*}[h!]
	\centering
	\includegraphics[width=1\linewidth]{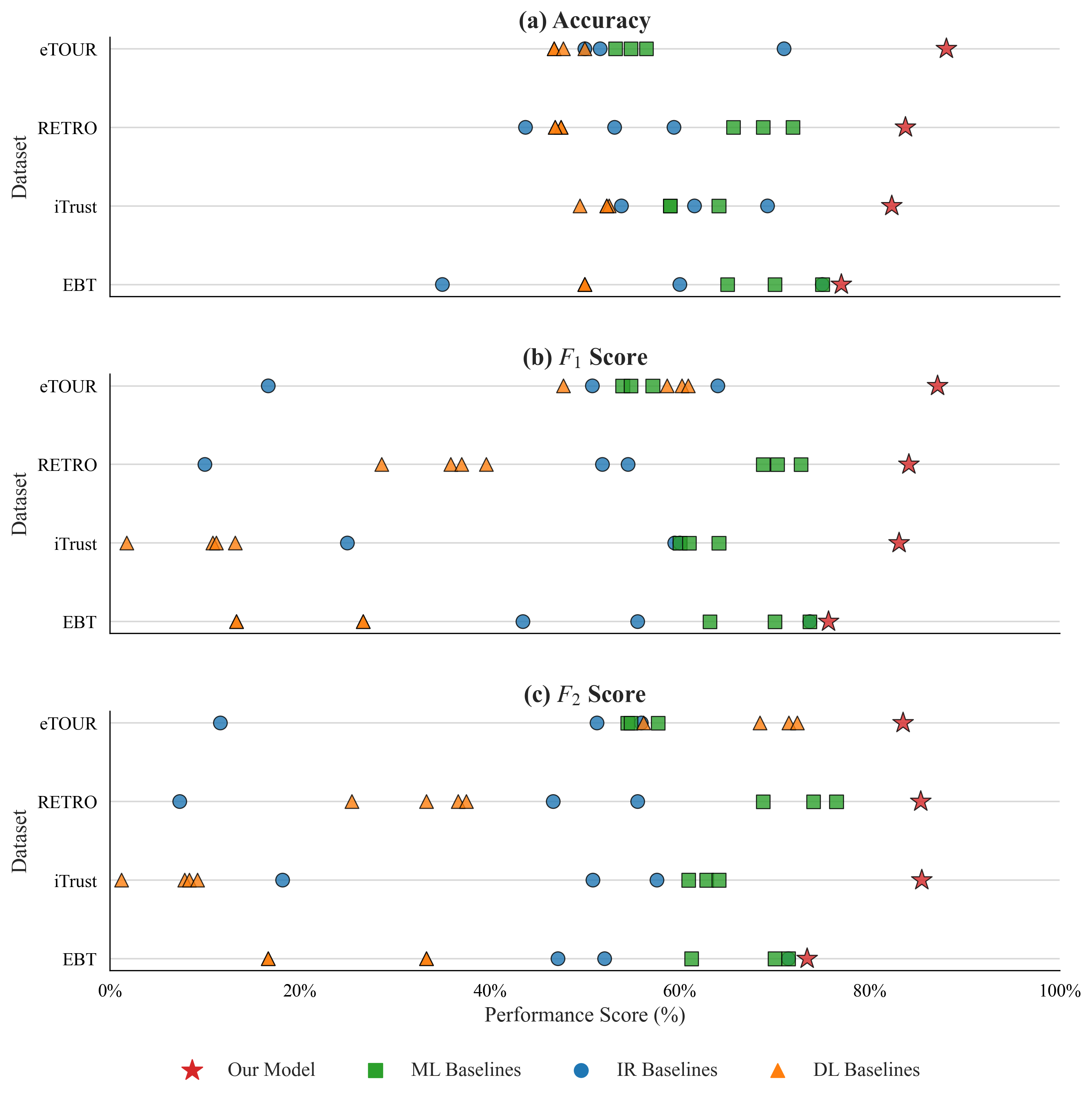}
	\caption{Benchmark Comparison Against Baseline Methods. The figure compares the performance of our proposed model against baselines from three dominant classes of techniques: information retrieval (IR), machine learning (ML), and deep learning (DL). Each subplot shows a different evaluation metric: (a) Accuracy, (b) $F_{1}$ Score, and (c) $F_{2}$ Score. Our model ($\bigstar$) consistently outperforms all baseline methods across all datasets, which are represented by shape: IR ($\bullet$), ML ($\blacksquare$), and DL ($\blacktriangle$).}
	\label{fig:baselines}
\end{figure*}

To this end, we conducted a comprehensive benchmark. For a fair and rigorous comparison, both our proposed model and the ten baselines were trained and evaluated on the identical augmented dataset used in Section~\ref{sec:rq3}—the one generated by GPT-4o employing the zero-shot requirement generation strategy. Our proposed model was specifically configured with the Jina encoder using a maximum sequence length of 512 tokens. The ten baselines represent the three dominant classes of techniques in the field: information retrieval, traditional machine learning, and deep learning, as detailed in Section~\ref{sec:baselines}. The results of this benchmark are presented in two complementary formats: Figure~\ref{fig:baselines} provides a visual comparison of our model against all baselines using dot plots for the three key metrics, while the precise numerical data is available for detailed inspection in~\ref{sec:comprehensive-results} ~(Tables~\ref{tab:baseline_1} and~\ref{tab:baseline_2}).

\textbf{\textit{Findings and Interpretation}}: Among the baselines, the supervised ML classifiers (e.g., SVM, LR) generally outperformed the classic IR methods. A striking observation was the poor performance of the conventional DL architectures (GRU, LSTM, and their variants), which often failed to surpass even the simpler IR approaches. This is likely attributable to their inherent data-hungriness; these models are trained from scratch and require vast amounts of data to learn complex semantic patterns effectively. In contrast, our framework leverages a pre-trained encoder, which has already acquired rich semantic knowledge from a massive pre-training corpus. This makes our approach far more data-efficient, whereas even our augmented dataset was likely insufficient for the DL baselines to converge to a robust solution.

Against this backdrop, the dot plots in Figure~\ref{fig:baselines} visually demonstrate our model's commanding performance advantage. Across all four datasets and all three key metrics, our model (represented by the star marker) is consistently positioned at the highest end of the performance spectrum, establishing a clear and substantial margin over the entire cluster of baseline methods. This comprehensive superiority is evidenced by the maximum performance gaps observed across the key metrics: our model outperformed the best-performing baseline by up to 17.09 percentage points on accuracy (eTOUR), 23.13 percentage points on $F_1$ score (eTOUR), and, most strikingly, by 21.37 percentage points on the recall-sensitive $F_2$ score (iTrust). This comprehensive benchmark, supported by both visual plots and detailed numerical data, provides compelling evidence that validates our central premise: the synergy between LLM-based data augmentation and an advanced, aligned pre-trained encoder creates a more powerful and robust solution for requirements traceability link recovery.




\textbf{\textit{Answer to RQ4}}: The benchmark evaluation confirms that our fully optimized framework conclusively outperforms all ten established baselines across the IR, ML, and DL techniques. This decisive advantage, particularly on the recall-sensitive $F_2$ score, demonstrates that the proposed synergistic approach effectively overcomes the data-scarcity bottleneck that limits traditional models, thereby setting a new, high-performance benchmark for requirements traceability link recovery on the evaluated datasets.


\section{Discussion}

This study addresses a critical and persistent bottleneck in requirements traceability: the scarcity of large-scale, high-quality labeled data. By systematically integrating LLM-driven data augmentation with an advanced, aligned pre-trained encoder, we have developed and validated a framework that largely enhances the performance of PLM-based traceability link recovery models. The central contribution of this work, therefore, is not merely an incremental improvement in model accuracy, but the establishment of a pragmatic and scalable pathway for applying advanced, data-driven requirements traceability solutions in industrial contexts where training data is inherently limited.

\subsection{Implications of Principal Findings}

Our experimental results, systematically organized around the four research questions, provide a multi-faceted validation of our proposed framework. The following discussion interprets these principal findings, moving beyond the numerical results to explore their deeper implications for the field of requirements traceability.

A primary contribution of this work stems from the comprehensive analysis of LLM-based data augmentation (RQ1 and RQ2). While our findings confirm the unequivocal effectiveness of this approach in enhancing model performance, they also reveal a more profound insight: for this task, the prompting strategy is the dominant factor, while the choice between leading LLMs is a secondary concern. This has significant practical implications. It suggests that practitioners and researchers can achieve substantial performance gains by focusing their efforts on the careful design of prompting strategies, rather than on the costly and often complex process of selecting or switching between different proprietary LLMs. This finding effectively decouples our data augmentation methodology from any single model provider, greatly enhancing its generalizability and practical applicability.

From a practical standpoint, these findings offer valuable guidance for the future application and scaling of this framework. Our results demonstrate that generating high-quality synthetic requirements from existing code is a highly effective prompting strategy for data augmentation. This is particularly promising for real-world scenarios where projects may possess an extensive and well-maintained codebase but lack correspondingly detailed requirement specifications~\citep{Medeiros2020,Franch2023}. In such contexts, the zero-shot requirement generation strategy stands out as not only highly effective—as proven by our experiment—but also as a pragmatic and scalable approach to enriching traceability datasets.

Our investigation into encoder optimization (RQ3) further refines our understanding of how to improve PLM-based method. The consistent and significant outperformance of the Jina encoder over CodeBERT, particularly on the C\#-based RETRO.NET dataset, provides strong empirical evidence that the degree of alignment between an encoder's pre-training corpus and the target programming languages is a critical performance driver. Furthermore, our analysis of input sequence length offers an important cautionary note: simply increasing the context window is not a universally effective strategy. Its dataset-dependent nature confirms that this should be treated as a sensitive hyperparameter, requiring empirical validation rather than being accepted as a default improvement.

Finally, the benchmark comparison against established techniques (RQ4) situates the value of our fully optimized framework. The commanding performance advantage of our model over all IR, ML, and traditional DL baselines demonstrates the superiority of our synergistic approach. More importantly, the stark contrast with the data-hungry DL baselines (e.g., GRU, LSTM) highlights a fundamental paradigm shift: in the data-scarce scenarios typical of requirements engineering, leveraging the rich, pre-existing knowledge embedded in advanced encoders is a far more effective and data-efficient strategy than attempting to learn complex semantic patterns from scratch. This finding validates our central premise that the synergy between data augmentation and a well-suited pre-trained encoder is key to overcoming the data scarcity bottleneck in requirements traceability.

\subsection{Relation to Prior Work}

Beyond interpreting our internal findings, it is crucial to situate this work within the broader landscape of requirements traceability research. Our framework does not operate in a vacuum; rather, it confirms, extends, and provides a novel solution to challenges well-documented in the literature.

The evolution of automated RTLR has followed a clear performance trajectory. Foundational studies~\citep{Guo2017,Dai2023,Wang2023a} established that deep learning models, such as RNNs and their variants (e.g., GRU, LSTM), offered superior semantic modeling capabilities and consistently outperformed traditional information retrieval methods like VSM and LSI, which were hampered by the vocabulary mismatch problem.

Subsequently, the advent of pre-trained language models marked another significant paradigm shift. The seminal work by Lin et al.~\citep{Lin2021} was pivotal in this transition, demonstrating that a fine-tuned BERT-based model decisively surpassed the previously dominant DL models. Their proposed ``Single" architecture, which concatenates the requirement and code into a single input sequence, is now regarded as an established state-of-the-art PLM-based traceability model. Our work is therefore positioned as a direct extension of this paradigm. We adopt the empirically validated ``Single" architecture as the foundation for our traceability classifier and focus on addressing the primary limitation that even this advanced approach faces: the scarcity of labeled training data~\citep{Zogaan2017,Lin2022}.

While our study confirms the superiority of the PLM-based architecture, its core contribution lies in proposing and validating a systematic, LLM-driven data augmentation framework as an effective remedy for data scarcity. The effectiveness of this framework is substantiated by our comprehensive evaluation, which benchmarks our enhanced framework not only against the established PLM baseline~\citep{Lin2021} but also against representative methods from all preceding paradigms, including classic IR methods (e.g., VSM, LSI)~\citep{Mahmoud2014,Rahimi2018}, traditional ML classifiers (e.g., SVM, LR)~\citep{Li2015,Mazrae2021}, and DL models (e.g., LSTM, GRU)~\citep{Guo2017}. By doing so, our research creates a direct ``echo" to the research gap we established in our introduction. We began by positing that the lack of large, high-quality datasets was the principal barrier to applying advanced ML models in industrial traceability scenarios; our discussion now closes this loop by demonstrating a pragmatic and powerful methodology to overcome that very barrier.

%

\subsection{Threats to Validity}


A primary threat to internal validity stems from the sequential nature of our multi-stage experimental design. Our framework was optimized in a stepwise manner: we first identified a top tier of prompting strategies using a baseline encoder (CodeBERT), and then used a representative strategy from that tier to evaluate and select a superior encoder (Jina). This approach, while pragmatic, assumes that the optimal prompting strategy is independent of the encoder architecture. It is conceivable that an alternative prompting strategy could have proven optimal when paired with the more advanced Jina encoder. A fully factorial experiment, while computationally prohibitive, would be required to definitively eliminate this threat. Therefore, a promising avenue for future work is to investigate the potential interaction effects between different data augmentation strategies and advanced encoder architectures.

The threats to external validity primarily concern the generalizability of our findings. First, our study is subject to a threat to temporal validity due to the rapid evolution of LLMs. While the LLMs used were state-of-the-art at the time of our experiments, the landscape of generative AI is constantly shifting. Although our core finding—that the prompting strategy is a more critical factor than the choice between leading LLMs—is likely to remain robust, the specific performance benchmarks we established may be surpassed by newer model generations. Future work should therefore focus on replicating these findings with subsequent generations of LLMs to assess the long-term stability of our conclusions. 

Second, our evaluation was conducted on four benchmark datasets encompassing three programming languages (Java, JavaScript, and C\#). While the results were consistent across these contexts, future research is needed to validate the framework's effectiveness on a broader range of programming languages, such as Python and C++, and on larger-scale industrial datasets. Second, our data augmentation relied exclusively on closed-source LLMs due to their current performance advantages~\citep{Hou2024}. A valuable next step would be to evaluate and incorporate leading open-source LLMs (e.g., Llama~\citep{Touvron2023} and DeepSeek~\citep{Yuan2025}) into our framework. This would not only test the approach's robustness with different model architectures but also enhance its accessibility and applicability for a wider range of users and organizations.

\section{Conclusion and Future Work}

This study addressed the critical bottleneck of data scarcity in requirements traceability by proposing and validating a synergistic framework that integrates LLM-driven data augmentation with an advanced, aligned encoder. Our research has yielded several key findings: we have demonstrated that LLM-based data augmentation is a highly effective strategy, with the choice of prompting strategy proving more critical than the specific LLM used, and that the performance of the traceability model can be further optimized by selecting an encoder whose pre-training corpus aligns with the target programming languages. The fully optimized framework demonstrates significant advancements on the evaluated datasets, conclusively outperforming a comprehensive suite of baselines and achieving performance gains of up to 28.59\% on $F_1$ score and 28.9\% on the recall-sensitive $F_2$ score over the established state-of-the-art PLM-based model. The principal contribution of this work, therefore, is the establishment of a pragmatic and scalable methodology for applying data-driven traceability solutions in data-scarce industrial contexts.

While this study validates a powerful new approach, its limitations define a clear roadmap for future research. A fully factorial experiment is needed to investigate the potential interaction effects between different data augmentation strategies and advanced encoder architectures. Furthermore, to enhance generalizability, the framework's effectiveness should be validated on a broader range of programming languages, such as Python and C++, and on large-scale industrial datasets. Future work should also replicate these findings with subsequent generations of LLMs to assess the long-term stability of our conclusions and explore the integration of leading open-source models to enhance the framework's accessibility. Beyond addressing these limitations, future research could also extend the application of this framework to other critical traceability tasks, such as linking requirements to test cases and design documents.



\clearpage

\appendix
\section{Data Augmentation Examples}
\label{sec:data-examples}

The following three boxes present a complete data augmentation example from the iTrust dataset. The synthetic artifacts were generated by the GPT-4o model using the few-shot prompt template.

\begin{tcolorbox}[
	title=A.1: Original Requirement-Code Pair,
	coltitle=black,
	colframe=promptboxframe,
	fonttitle=\bfseries,
	arc=3mm,
	fontupper=\small
	]
	\textcolor{promptheading}{\textbf{Original Requirement}} \\
	The patient, UAP, or personal representative enters a systolic blood pressure outside the range 40-240 or a diastolic blood pressure outside the range 40-150. He/she is notified of an error and is prompted to try again.
	\vspace{2mm}\hrule\vspace{2mm}
	
	\textcolor{promptheading}{\textbf{Original Code}}
\begin{lstlisting}
package edu.ncsu.csc.itrust.validate;

import edu.ncsu.csc.itrust.action.AddRemoteMonitoringDataAction;
import edu.ncsu.csc.itrust.beans.RemoteMonitoringDataBean;
import edu.ncsu.csc.itrust.exception.ErrorList;
import edu.ncsu.csc.itrust.exception.FormValidationException;

public class RemoteMonitoringDataBeanValidator extends BeanValidator<RemoteMonitoringDataBean> {
	public RemoteMonitoringDataBeanValidator() {
	}
	
	@Override
	public void validate(RemoteMonitoringDataBean m) throws FormValidationException {
		ErrorList errorList = new ErrorList();
		errorList.addIfNotNull(checkFormat("Systolic Blood Pressure", "" + m.getSystolicBloodPressure(),
		ValidationFormat.SYSTOLIC_BLOOD_PRESSURE, false));
		errorList.addIfNotNull(checkFormat("Diastolic Blood Pressure", "" + m.getDiastolicBloodPressure(),
		ValidationFormat.DIASTOLIC_BLOOD_PRESSURE, false));
		errorList.addIfNotNull(checkFormat("Glucose Level", "" + m.getGlucoseLevel(),
		ValidationFormat.GLUCOSE_LEVEL, false));
		if (errorList.hasErrors()){
			throw new FormValidationException(errorList);
		}
	}
}	
\end{lstlisting}
\end{tcolorbox}

\begin{tcolorbox}[
	breakable,
	title=A.2: R-to-C Synthetic Pair (Requirement \(\to\) Code),
	coltitle=black,
	colframe=promptboxframe,
	fonttitle=\bfseries,
	arc=3mm,
	fontupper=\small
	]
	\textcolor{promptheading}{\textbf{Original Requirement}} \\
	The patient, UAP, or personal representative enters a systolic blood pressure outside the range 40-240 or a diastolic blood pressure outside the range 40-150. He/she is notified of an error and is prompted to try again.
	\vspace{2mm}\hrule\vspace{2mm}
	
	\textcolor{promptheading}{\textbf{Synthetic Code}}
\begin{lstlisting}
package edu.ncsu.csc.itrust.action;

import edu.ncsu.csc.itrust.exception.FormValidationException;
import edu.ncsu.csc.itrust.beans.BloodPressureBean;

public class BloodPressureAction {
	public void validateBloodPressure(int systolic, int diastolic) throws FormValidationException {
		if (systolic < 40 || systolic > 240) {
			throw new FormValidationException("Systolic blood pressure must be between 40 and 240.");
		}
		if (diastolic < 40 || diastolic > 150) {
			throw new FormValidationException("Diastolic blood pressure must be between 40 and 150.");
		}
	}
	public boolean setBloodPressure(BloodPressureBean bpBean) {
		try {
			validateBloodPressure(bpBean.getSystolic(), bpBean.getDiastolic());
			return true;
		} catch (FormValidationException e) {
			System.out.println("Error: " + e.getMessage());
			System.out.println("Please try again.");
			return false;
		}
	}
}

package edu.ncsu.csc.itrust.beans;

public class BloodPressureBean {
	private int systolic;
	private int diastolic;
	
	public int getSystolic() {
		return systolic;
	}
	public void setSystolic(int systolic) {
		this.systolic = systolic;
	}
	public int getDiastolic() {
		return diastolic;
	}
	public void setDiastolic(int diastolic) {
		this.diastolic = diastolic;
	}
}

package edu.ncsu.csc.itrust.exception;

public class FormValidationException extends Exception {
	private static final long serialVersionUID = 1L;
	
	public FormValidationException(String message) {
		super(message);
	}
}
\end{lstlisting}
\end{tcolorbox}

\begin{tcolorbox}[
	title=A.3: C-to-R Synthetic Pair (Code \(\to\) Requirement),
	coltitle=black,
	colframe=promptboxframe,
	fonttitle=\bfseries,
	arc=3mm,
	fontupper=\small
	]
	\textcolor{promptheading}{\textbf{Synthetic Requirement}} \\
	1. Users expect to validate remote monitoring data.
	2. Users are required to provide valid systolic blood pressure, diastolic blood pressure, and glucose level measurements when adding remote monitoring data.
	3. If the provided remote monitoring data does not meet the specified formats for systolic blood pressure, diastolic blood pressure, or glucose level, the user will receive validation errors indicating the incorrect fields.
	\vspace{2mm}\hrule\vspace{2mm}
	
	\textcolor{promptheading}{\textbf{Original Code}}
\begin{lstlisting}
package edu.ncsu.csc.itrust.validate;

import edu.ncsu.csc.itrust.action.AddRemoteMonitoringDataAction;
import edu.ncsu.csc.itrust.beans.RemoteMonitoringDataBean;
import edu.ncsu.csc.itrust.exception.ErrorList;
import edu.ncsu.csc.itrust.exception.FormValidationException;

public class RemoteMonitoringDataBeanValidator extends BeanValidator<RemoteMonitoringDataBean> {
	public RemoteMonitoringDataBeanValidator() {
	}
	
	@Override
	public void validate(RemoteMonitoringDataBean m) throws FormValidationException {
		ErrorList errorList = new ErrorList();
		errorList.addIfNotNull(checkFormat("Systolic Blood Pressure", "" + m.getSystolicBloodPressure(),
		ValidationFormat.SYSTOLIC_BLOOD_PRESSURE, false));
		errorList.addIfNotNull(checkFormat("Diastolic Blood Pressure", "" + m.getDiastolicBloodPressure(),
		ValidationFormat.DIASTOLIC_BLOOD_PRESSURE, false));
		errorList.addIfNotNull(checkFormat("Glucose Level", "" + m.getGlucoseLevel(),
		ValidationFormat.GLUCOSE_LEVEL, false));
		if (errorList.hasErrors()){
			throw new FormValidationException(errorList);
		}
	}
}
\end{lstlisting}
\end{tcolorbox}

\section{The comprehensive TRLR results}
\label{sec:comprehensive-results}

\begin{table}[!ht]
	\caption{Performance comparison of each combination of LLM and prompting strategy on the EBT and eTOUR datasets. The best-performing average result for each metric is shown in \textbf{bold}.}
	\centering
	\begin{tabular}{@{} c| c| c c c c c c c @{}}
		
		\toprule
		\textbf{Dataset} & \textbf{Prompting Strategy} & \textbf{Metric} & \textbf{Claude3} & \textbf{Gemini} & \textbf{GPT-3.5} & \textbf{GPT-4o} & \textbf{Average} & \textbf{Std.} \\
		
		\midrule  
		
		
		\multirow{10}{*}{\textbf{EBT}}
		& \multirow{2}{*}{None} & F1 & - & - & - & - & 65.93\% & - \\ 
		~ & ~ & F2 & - & - & - & - & 63.49\% & - \\
		\cline{3-9}
		~ & \multirow{2}{*}{ZS-Code} & F1 & 65.93\% & 70.00\% & 67.13\% & 67.33\% & \textbf{67.60\%} & 1.72\% \\ 
		~ & ~ & F2 & 64.72\% & 68.73\% & 65.11\% & 65.28\% & 65.96\% & 1.86\% \\ 
		\cline{3-9}
		~ & \multirow{2}{*}{ZS-Req} & F1 & 66.99\% & 66.70\% & 65.73\% & 69.45\% & 67.22\% & 1.58\% \\ 
		~ & ~ & F2 & 65.00\% & 68.52\% & 63.38\% & 68.49\% & \textbf{66.35\%} & 2.58\% \\ 
		\cline{3-9}
		~ & \multirow{2}{*}{FS-Code} & F1 & 66.12\% & 66.63\% & 63.96\% & 65.96\% & 65.67\% & 1.18\% \\ 
		~ & ~ & F2 & 63.54\% & 63.74\% & 63.92\% & 65.92\% & 64.28\% & 1.10\% \\ 
		\cline{3-9}
		~ & \multirow{2}{*}{FS-Req} & F1 & 67.59\% & 71.63\% & 67.12\% & 62.70\% & 67.26\% & 3.65\% \\ 
		~ & ~ & F2 & 65.24\% & 69.37\% & 62.56\% & 60.79\% & 64.49\% & 3.73\% \\

		\hline
		
		\multirow{10}{*}{\textbf{eTOUR}}
		& \multirow{2}{*}{None} & F1 & - & - & - & - & 58.54\% & - \\ 
		~ & ~ & F2 & - & - & - & - & 54.59\% & - \\ 
		\cline{3-9}
		~ & \multirow{2}{*}{ZS-Code} & F1 & 85.03\% & 82.01\% & 83.81\% & 83.14\% & \textbf{83.50\%} & 1.26\% \\ 
		~ & ~ & F2 & 82.33\% & 80.38\% & 81.48\% & 80.80\% & \textbf{81.25\%} & 0.85\% \\ 
		\cline{3-9}
		~ & \multirow{2}{*}{ZS-Req} & F1 & 83.35\% & 86.32\% & 80.66\% & 81.12\% & 82.86\% & 2.59\% \\ 
		~ & ~ & F2 & 79.68\% & 82.78\% & 77.83\% & 78.84\% & 79.78\% & 2.14\% \\ 
		\cline{3-9}
		~ & \multirow{2}{*}{FS-Code} & F1 & 84.73\% & 85.51\% & 80.86\% & 80.47\% & 82.89\% & 2.60\% \\ 
		~ & ~ & F2 & 80.94\% & 82.90\% & 77.52\% & 78.59\% & 79.99\% & 2.41\% \\ 
		\cline{3-9}
		~ & \multirow{2}{*}{FS-Req} & F1 & 67.12\% & 77.37\% & 83.17\% & 80.44\% & 77.02\% & 7.02\% \\ 
		~ & ~ & F2 & 66.64\% & 77.00\% & 82.75\% & 78.98\% & 76.34\% & 6.89\% \\

		\bottomrule
	\end{tabular}
	\label{tbl:combination1}
\end{table}

\begin{table}[!ht]
	
	\centering
	\caption{Performance comparison of each combination of LLM and prompting strategy on the iTrust and RETRO datasets. The best-performing average result for each metric is shown in \textbf{bold}.}
	\begin{tabular}{@{} c| c| c c c c c c c @{}}
		
		\toprule
		\textbf{Dataset} & \textbf{Prompting Strategy} & \textbf{Metric} & \textbf{Claude3} & \textbf{Gemini} & \textbf{GPT-3.5} & \textbf{GPT-4o} & \textbf{Average} & \textbf{Std.} \\
		
		\midrule  
		
		\multirow{10}{*}{\textbf{iTrust}}
		& \multirow{2}{*}{None} & F1 & - & - & - & - & 67.19\% & - \\ 
		~ & ~ & F2 & - & - & - & - & 69.06\% & - \\ 
		\cline{3-9}
		~ & \multirow{2}{*}{ZS-Code} & F1 & 83.25\% & 86.40\% & 84.23\% & 84.45\% & \textbf{84.58\%} & 1.32\% \\ 
		~ & ~ & F2 & 84.65\% & 89.24\% & 86.55\% & 86.36\% & 86.70\% & 1.90\% \\ 
		\cline{3-9}
		~ & \multirow{2}{*}{ZS-Req} & F1 & 83.11\% & 85.08\% & 79.54\% & 79.61\% & 81.83\% & 2.73\% \\ 
		~ & ~ & F2 & 87.50\% & 88.08\% & 82.20\% & 81.29\% & 84.77\% & 3.52\% \\ 
		\cline{3-9}
		~ & \multirow{2}{*}{FS-Code} & F1 & 83.96\% & 86.34\% & 83.40\% & 83.85\% & 84.39\% & 1.32\% \\ 
		~ & ~ & F2 & 89.05\% & 90.67\% & 87.36\% & 86.99\% & \textbf{88.52\%} & 1.69\% \\ 
		\cline{3-9}
		~ & \multirow{2}{*}{FS-Req} & F1 & 76.22\% & 78.03\% & 83.06\% & 83.88\% & 80.30\% & 3.75\% \\ 
		~ & ~ & F2 & 79.27\% & 79.18\% & 86.65\% & 86.99\% & 83.02\% & 4.39\% \\ 
		
		\hline
		
		\multirow{10}{*}{\textbf{RETRO}}
		& \multirow{2}{*}{None} & F1 & - & - & - & - & 68.38\% & - \\ 
		~ & ~ & F2 & - & - & - & - & 66.27\% & - \\ 
		\cline{3-9}
		~ & \multirow{2}{*}{ZS-Code} & F1 & 79.72\% & 81.55\% & 79.68\% & 81.73\% & \textbf{80.67\%} & 1.12\% \\ 
		~ & ~ & F2 & 79.13\% & 82.10\% & 79.85\% & 82.92\% & \textbf{81.00\%} & 1.80\% \\ 
		\cline{3-9}
		~ & \multirow{2}{*}{ZS-Req} & F1 & 72.39\% & 67.50\% & 74.14\% & 73.42\% & 71.86\% & 3.00\% \\ 
		~ & ~ & F2 & 70.89\% & 66.70\% & 73.84\% & 74.34\% & 71.44\% & 3.51\% \\ 
		\cline{3-9}
		~ & \multirow{2}{*}{FS-Code} & F1 & 74.64\% & 75.18\% & 68.28\% & 74.70\% & 73.20\% & 3.29\% \\ 
		~ & ~ & F2 & 74.74\% & 74.30\% & 66.26\% & 74.06\% & 72.34\% & 4.06\% \\ 
		\cline{3-9}
		~ & \multirow{2}{*}{FS-Req} & F1 & 56.94\% & 70.36\% & 63.66\% & 64.70\% & 63.91\% & 5.50\% \\ 
		~ & ~ & F2 & 56.30\% & 70.87\% & 62.90\% & 63.26\% & 63.33\% & 5.96\% \\ 
		\bottomrule
	\end{tabular}
	\label{tbl:combination2}
\end{table}

\begin{table}[!ht]
	\caption{Performance comparison of our model against Information Retrieval (IR), Machine Learning (ML), and Deep Learning (DL) baselines on the EBT and eTOUR datasets. The best-performing result for each metric is shown in \textbf{bold}, and the second-highest result is \underline{underlined}.}
	
	\centering
	\begin{tabular}{lccc|ccc}
			\toprule
			\multicolumn{1}{l}{} & \multicolumn{3}{c}{EBT} & \multicolumn{3}{c}{eTOUR} \\
			\multicolumn{1}{l}{} & Accuracy & $F_{1}$ & $F_{2}$ & Accuracy & $F_{1}$ & $F_{2}$ \\
			\midrule
			LDA & 35.00\% & 43.48\% & 47.17\% & 51.61\% & 16.67\% & 11.63\% \\
			LSI & \underline{75.00\%} & \underline{73.68\%} & \underline{71.43\%} & 50.00\% & 50.79\% & 51.28\% \\
			VSM & 60.00\% & 55.56\% & 52.08\% & \underline{70.97\%} & \underline{64.00\%} & 55.94\% \\
			KNN & 65.00\% & 63.16\% & 61.22\% & 53.23\% & 53.97\% & 54.49\% \\
			LR & 70.00\% & 70.00\% & 70.00\% & 56.45\% & 57.14\% & 57.69\% \\
			SVM & \underline{75.00\%} & \underline{73.68\%} & \underline{71.43\%} & 54.84\% & 54.84\% & 54.84\% \\
			GRU & 50.00\% & 13.33\% & 16.67\% & 46.77\% & 60.23\% & 71.48\% \\
			LSTM & 50.00\% & 26.67\% & 33.33\% & 50.00\% & 47.75\% & 56.19\% \\
			Bi-GRU & 50.00\% & 13.33\% & 16.67\% & 46.77\% & 58.67\% & 68.44\% \\
			Bi-LSTM & 50.00\% & 26.67\% & 33.33\% & 47.74\% & 60.89\% & \underline{72.37\%} \\
			\hline
			Ours & \textbf{77.00\%} & \textbf{75.65\%} & \textbf{73.40\%} & \textbf{88.06\%} & \textbf{87.13\%} & \textbf{83.50\%} \\
			\bottomrule
		\end{tabular}
		\label{tab:baseline_1}
\end{table}

\begin{table}[!ht]
	\caption{Performance comparison of our model against Information Retrieval (IR), Machine Learning (ML), and Deep Learning (DL) baselines on the iTrust and RETRO datasets. The best-performing result for each metric is shown in \textbf{bold}, and the second-highest result is \underline{underlined}.}
	
	\centering
	\begin{tabular}{lccc|ccc}
			\toprule
			\multicolumn{1}{l}{} & \multicolumn{3}{c}{iTrust} & \multicolumn{3}{c}{RETRO} \\
			\multicolumn{1}{l}{} & Accuracy & $F_{1}$ & $F_{2}$ & Accuracy & $F_{1}$ & $F_{2}$ \\
			\midrule
			LDA & 53.85\% & 25.00\% & 18.18\% & 43.75\% & 10.00\% & 7.35\% \\
			LSI & 61.54\% & 59.46\% & 57.59\% & 53.13\% & 54.55\% & 55.56\% \\
			VSM & \underline{69.23\%} & 60.00\% & 50.85\% & 59.38\% & 51.85\% & 46.67\% \\
			KNN & 58.97\% & 60.00\% & 60.91\% & 65.63\% & 70.27\% & \underline{76.47\%} \\
			LR & 58.97\% & 60.98\% & 62.81\% & \underline{71.88\%} & \underline{72.73\%} & 74.07\% \\
			SVM & 64.10\% & \underline{64.10\%} & \underline{64.10\%} & 68.75\% & 68.75\% & 68.75\% \\
			GRU & 52.31\% & 10.84\% & 7.88\% & 47.50\% & 39.63\% & 36.67\% \\
			LSTM & 52.56\% & 11.21\% & 8.37\% & 47.50\% & 35.89\% & 37.53\% \\
			Bi-GRU & 49.49\% & 1.78\% & 1.23\% & 46.88\% & 37.04\% & 33.33\% \\
			Bi-LSTM & 52.31\% & 13.20\% & 9.23\% & 46.88\% & 28.62\% & 25.48\% \\
			\hline
			Ours & \textbf{82.31\%} & \textbf{83.08\%} & \textbf{85.47\%} & \textbf{83.75\%} & \textbf{84.11\%} & \textbf{85.36\%} \\
			\bottomrule
		\end{tabular}
		\label{tab:baseline_2}
\end{table}


\clearpage

\bibliographystyle{elsarticle-harv}
\bibliography{cii-ref.bib}



%
%
%
\end{document}